\documentclass[acmtog]{acmart}

\acmSubmissionID{500}

\hypersetup{draft}

\usepackage{booktabs} 

\setcitestyle{square}

\usepackage[ruled]{algorithm2e} 

\SetAlFnt{\small}
\SetAlCapFnt{\small}
\SetAlCapNameFnt{\small}
\SetAlCapHSkip{0pt}
\IncMargin{-\parindent}

\acmJournal{TOG}

\setcopyright{none}

\acmJournal{TOG}
\acmYear{2020}\acmVolume{39}\acmNumber{4}\acmArticle{1}\acmMonth{7} \acmDOI{10.1145/3386569.3392469}

\newcommand{\etal}{et al.}
\usepackage{color}
\usepackage{soul}
\usepackage{breqn}
\usepackage{booktabs}
\usepackage{multirow}
\usepackage{rotating}

\definecolor{purple}{rgb}{0.65,0,0.65}
\definecolor{blue}{rgb}{0, 0.2, 0.8}
\definecolor{orange}{rgb}{0.6, 0.6, 0}
\definecolor{red}{rgb}{0.8, 0.2, 0.2}
\definecolor{magenta}{rgb}{0.5, 0.0, 1.0}
\definecolor{black}{rgb}{0.0, 0.0, 0.0}
\definecolor{cyan}{rgb}{0, 0.65, 0.65}
\definecolor{olive}{rgb}{0.2, 0.6, 0.5}
\usepackage{xcolor}

\newif\ifdraft
\drafttrue

\ifdraft
\newcommand{\dcc}[1]{{\color{red}\textbf{DC:} #1}}
\newcommand{\kac}[1]{{\color{orange}\textbf{KA:} #1}}
\newcommand{\dlc}[1]{{\color{magenta}\textbf{DL:} #1}}
\newcommand{\ywc}[1]{{\color{green}\textbf{YW:} #1}}

\else
\newcommand{\dlc}[1]{}
\newcommand{\dcc}[1]{}
\newcommand{\kac}[1]{}
\newcommand{\ywc}[1]{}

\fi

\newcommand{\bm}{{\bf m}}
\newcommand{\bn}{{\bf n}}

\newcommand{\bw}{{\bf w}}
\newcommand{\bx}{{\bf x}}
\newcommand{\by}{{\bf y}}
\newcommand{\bz}{{\bf z}}

\newcommand{\Loss}{\mathcal{L}}
\newcommand{\mm}{\mathcal{M}}
\newcommand{\mms}{\mathcal{S}}

\newcommand{\bbe}{\mathbb{E}}
\def \figures {./}

\begin{document}

\title{Unpaired Motion Style Transfer from Video to Animation}

\author{Kfir Aberman}
\affiliation{
	\institution{AICFVE, Bejing Film Academy \& Tel-Aviv University}
}
\authornote{equal contribution}

\author{Yijia Weng}
\affiliation{
	\institution{CFCS, Peking University \& AICFVE, Beijing Film Academy}
}
\authornotemark[1]

\author{Dani Lischinski}
\affiliation{
	\institution{The Hebrew University of Jerusalem \& AICFVE, Beijing Film Academy}
}

\author{Daniel Cohen-Or}
\affiliation{
	\institution{Tel-Aviv University \& AICFVE, Beijing Film Academy}
}

\author{Baoquan Chen}
\affiliation{
	\institution{CFCS, Peking University \& AICFVE, Beijing Film Academy}
}
\authornote{corresponding author}

\renewcommand\shortauthors{Aberman, K. et al.}

\authorsaddresses{%
Authors' addresses: Kfir Aberman, kfiraberman@gmail.com; Yijia Weng, halfsummer11@gmail.com; Dani Lischinski, danix3d@gmail.com; Daniel Cohen-Or, cohenor@gmail.com; Baoquan Chen, \mbox{baoquan@pku.edu.cn}}

\begin{abstract}
Transferring the motion style from one animation clip to another, while preserving the motion content of the latter, has been a long-standing problem in character animation. 
Most existing data-driven approaches are supervised and rely on paired data, where motions with the same content are performed in different styles.
In addition, these approaches are limited to transfer of styles that were seen during training.

In this paper, we present a novel data-driven framework for motion style transfer, which learns from an unpaired collection of motions with style labels, and enables transferring motion styles not observed during training.
Furthermore, our framework is able to extract motion styles directly from videos, bypassing 3D reconstruction, and apply them to the 3D input motion. 

Our style transfer network encodes motions into two latent codes, for content and for style, each of which plays a different role in the decoding (synthesis) process. While the content code is decoded into the output motion by several temporal convolutional layers, the style code modifies deep features via temporally invariant adaptive instance normalization (AdaIN). 

Moreover, while the content code is encoded from 3D joint rotations, we learn a common embedding for style from either 3D or 2D joint positions, enabling style extraction from videos.

Our results are comparable to the state-of-the-art, despite not requiring paired training data, and outperform other methods when transferring previously unseen styles. To our knowledge, we are the first to demonstrate style transfer directly from videos to 3D animations - an ability which enables one to extend the set of style examples far beyond motions captured by MoCap systems.

\end{abstract}

%

%
%




\maketitle

\begin{figure}
\centering
\includegraphics[width=\linewidth]{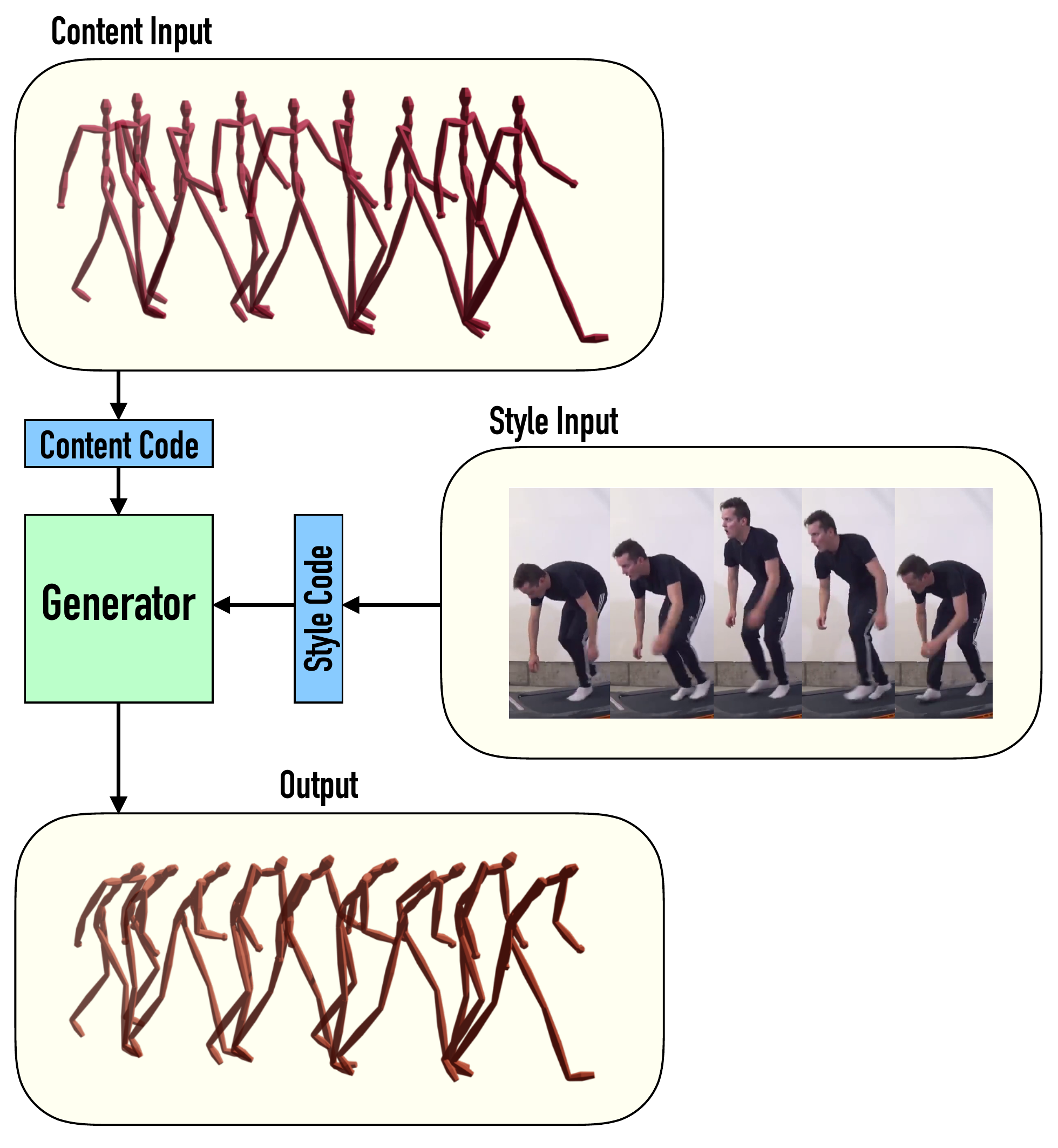} 
\caption{Style transfer from video to animation. Our network, which is trained with unpaired motion sequences, learns to disentangle content and style. Our trained generator is able to produce a motion sequence that combines the content of a 3D sequence with the style extracted directly from a video.} 
\label{fig:teaser}
\end{figure}
\section{Introduction}
\label{sec:intro}

The style of human motion may be thought of as the collection of motion attributes that convey the mood and the personality of a character. Human observers are extremely perceptive to subtle style variations; we can, for example, often tell whether a person is happy or sad from the way they walk. Consequently, for games and movies that pursue realistic and expressive character animation, there is a long-standing interest in generating diverse stylized motions. However, capturing all desired motions in a variety of styles is practically infeasible. A much more promising option is to perform motion style transfer: modify the style of an existing motion into one taken from another. Furthermore, it is particularly attractive to use video clips to specify target motion styles. 

Since motion style eludes a precise definition, hand crafted representations are not well-suited to cope with style transfer, and most recent works attempt to infer style from examples.
However, despite years of progress in data-driven motion style transfer, two main obstacles remain in practice: (i) avoiding the need for paired and registered data, and (ii) extracting style from only a few examples. Both of these hurdles arise because of the difficulty to collect and process sufficient motion data.
In order to capture paired and registered data, the same actor must, for example, perform a walking sequence in different styles with identical steps and turns, which is tedious and, more importantly, unscalable to the huge training sets required by today's deep learning models.
As for the extraction of styles, clearly, a large number of style examples might better characterize the style and facilitate the transfer. However, in reality, one can often obtain only a few examples for each (uncommon) style.

Additionally, we also see great potential in videos as a massive source for motion styles. As the primary format for recording human activities, available videos contain a much wider range of motions and styles compared to 3D motion capture data. And if the desired style needs to be captured on-the-spot, shooting a video is a much easier and cheaper alternative to performing motion capture. 

As an important research topic, the problem of motion style transfer has been approached in several ways, all of which we find limited, especially with regard to the challenges above. Some works model style using hand-crafted representations, such as physical parameters or spectral characteristics, which may fail to fully capture complex and subtle properties. Other works adopt a data-driven approach. Their reliance on paired training data or large numbers of target style examples, however, hinders their applicability to real world settings. Recently, Mason et al.~\shortcite{mason2018few} proposed a few-shot learning scheme to cope with the shortage of style references. However, they employ a specialized network that targets only locomotion. Also, like other previous works, their model can only extract style from 3D MoCap data. 

In this work, we circumvent the need for paired training data by learning without supervision, characterize unseen styles even from a single reference clip, and provide the ability to extract style from a video. To achieve these goals, we adopt a generative scheme, using temporal convolutional neural networks as the backbone of our model. Our network encodes content and style inputs into corresponding latent codes, which are then recombined and decoded to obtain a re-stylized result (see Figure \ref{fig:teaser}). 
We argue that during training, our network learns a universal style extractor, which is then applicable to new styles, specified via a few examples at test time. Furthermore, our style extractor is applicable to both 3D motions and 2D motions observed in ordinary video examples, without requiring 3D reconstruction.

Our main technical contribution lies in the architecture of the deep neural network outlined above, in which  the content and style codes affect the generated motions via two different mechanisms. The content code is decoded into a motion by applying a sequence of temporal convolutional layers, each resulting in a set of temporal signals that represent joint rotations in a high-dimensional feature space. The style code is used to modify the second order statistics of these generated deep features via temporally-invariant adaptive instance normalization (AdaIN). These temporally-invariant affine transformations amplify or attenuate the temporal signals, while preserving their shape. Consequently, the motion content is also preserved. AdaIN has been used with great effect in image generation and style transfer (see Section~\ref{sec:rel_image_style_tran}), but to our knowledge we are the first to apply it in the context of motion.

Our network is trained by optimizing a \emph{content consistency loss}, which ensures that the content input to our network is reconstructed whenever the style input has the same style label as the content one. This loss forces the network to extract only those attributes that are shared among samples of the same style class. Simply copying the content input to the output is prevented by instance normalization during the encoding and by restricting the dimensionality of the latent content code.  

In order to extract style from videos we learn a joint embedding space for style codes extracted from either 3D or 2D joint positions. During training, we require that 3D motions and their 2D projections, are both mapped by a pair of corresponding encoders into the same style code. In addition to enabling style extraction from videos, this joint embedding supports style interpolation, as well as measuring ``style distance'' between videos and/or 3D motions. 

In summary, our contributions consist of a novel data-driven approach for motion style transfer that: (i) does not require paired training data; (ii) is able to transfer styles unseen during training, extracted from as little as a single example clip; and (iii) supports style extraction directly from ordinary videos.

We discuss various insights related to the mechanism of the network and the analogy to some related works. Our results show that by leveraging the power of our new framework, we can match state-of-the-art motion style transfer results, by training only on unpaired motion clips with style labels. Furthermore, we outperform other methods for previously unseen styles, as well as styles extracted from ordinary videos.

\section{Related Work}
\label{sec:rel}

\subsection{Image Style Transfer}
\label{sec:rel_image_style_tran}
Our work is inspired by the impressive progress in image style transfer, achieved through the use of deep learning machinery. The pioneering work of Gatys~\etal~\shortcite{gatys2016image} showed that style and content can be represented by statistics of deep features extracted from a pre-trained classification network. While their original approach required optimization to transfer style between each pair of images, Johnson~\etal~\shortcite{johnson2016perceptual} later converted this approach to a feed-forward one by training a network using a perceptual loss.

Later on, Ulyanov~\etal~\shortcite{ulyanov2016instance} showed that the style of an image can be manipulated by modifying the second order statistics (mean and variance) of channels of intermediate layers and proposed an instance normalization layer which enables to train a network to modify the style of arbitrary content images into a single specific target style. This idea was further extended by Huang~\etal~\shortcite{huang2017arbitrary}, who demonstrated that various target styles may be applied simply by using the Adaptive Instance Normalization (AdaIN) layer to inject different style statistics into the same network.

The AdaIN mechanism has proved effective for various tasks on images, such as image-to-image translation \cite{huang2018multimodal} and image generation \cite{karras2019style}. Recently, Park~\etal~\shortcite{park2019semantic} proposed a spatially adaptive normalization layer, for multi-modal generation of images, based on semantic segmentation maps, while Liu~\etal~\shortcite{liu2019few} introduced FUNIT, a few-shot unpaired image-to-image translation method, where only a few examples of the target class are required.

Inspired by these recent achievements in image style transfer, our work makes use of temporally invariant AdaIN parameters, thus enabling manipulating a given motion sequence to perform in arbitrary styles.
To our knowledge, we are the first to employ such a mechanism in the context of motion processing.

\subsection{Motion style transfer}

Motion style transfer is a long standing problem in computer animation.  Previous works relied on handcrafted features, in frequency domain \cite{unuma1995fourier}, or in time \cite{amaya1996emotion}, to represent and manipulate style or emotion \cite{aristidou2017emotion} of given 3D motions, or used physics-based optimizations \cite{liu2005learning} to achieve a similar goal. Yumer and Mitra~\shortcite{yumer2016spectral} showed that difference in spectral intensities of two motion signals with similar content but different styles enables the transfer between these two styles on arbitrary heterogeneous actions.

Since style is an elusive attribute that defies a precise mathematical definition, data-driven approaches that infer style features from examples, might have an advantage compared to attempting to hand-craft features to characterize style and content. 
Indeed, several works used machine learning tools to perform motion style transfer \cite{brand2000style,hsu2005style, wang2007multifactor, ikemoto2009generalizing, ma2010modeling,xia2015realtime}. However, these methods either assume that there are explicit motion pairs in the training data that exhibit the same motion with different styles, or limited to the set of styles given in the dataset.

For example, Hsu~\etal~\shortcite{hsu2005style} learned a linear translation model that can map a motion to a target style based on pairwise dense correspondence (per frame) between motions with similar content but different styles.  Xia~\etal~\shortcite{xia2015realtime} used a KNN search over a database of motions to construct a mixture of regression models for transferring style between motion clips, and Smith~\etal~\shortcite{smith2019efficient} improved the processing using a neural network that is trained on paired and registered examples.
Both latter methods represent style by a vector of fixed length, which is limited to the set of styles in the dataset, and can not be applied to unseen styles. In contrast, our approach only assumes that each motion clip is labeled by its style, without any requirement for content labels or correspondences between motion clips. In addition, our method enables extraction of unseen styles from  3D motion examples, and even from video, and is not limited to the styles in the dataset.

Besides transferring motion style, other methods exploited machine learning approaches to cope with various closely related tasks. These include generating motion with constraints using inverse kinematics (IK) when the style is taken from a dataset \cite{grochow2004style}, performing independent component analysis to separate motion into different components and perform a transfer \cite{shapiro2006style}, or using restricted Boltzmann machines, conditioned on a style label, to model human motion and capture style \cite{taylor2009factored}.

With the recent rapid progress in deep learning methods for character animation \cite{holden2015learning, holden2016deep,holden2017phase}, the flourishing image style transfer techniques were quickly adopted into the character animation domain. Holden~\etal~\shortcite{holden2016deep}, proposed a general learning framework for motion editing that enables style transfer. In analogy to Gatys~\etal~\shortcite{gatys2016image}, they optimize a motion sequence that satisfy two conditions; the activations of the hidden units should be similar to those of the  content motion, while their Gram matrices should match those of the style input motion. 
Differently from Gatys~\etal~\shortcite{gatys2016image}, here the features are extracted using a pretrained autoencoder for motion, rather than a pretrained image classification network. 
Later on, Holden~\etal~\shortcite{holden2017fast} and Du~\etal~\shortcite{du2019stylistic} proposed to improve performance by replacing optimization with a feed-forward network that is trained to satisfy the same constraints. In both of these works, the pretrained network is not explicitly designed for style transfer. The features extracted by the pretrained autoencoder contain information on both content and style, which leads to a strong dependency between the two properties, as demonstrated in Section~\ref{sec:exp}. 

Recently, Mason~\etal~\shortcite{mason2018few} proposed a method to transfer the style of character locomotion in real time, given a few shots of another stylized animation. In contrast to this approach, which is limited to locomotion, our approach can extract and transfer styles regardless of the motion content.

\subsection{Motion from Videos}
Motion reconstruction and pose estimation from monocular videos are long-standing fundamental tasks in computer vision, and are beyond the scope this paper.
Along the years, various methods have dealt with the extraction of different motion properties directly from video, such as 2D poses~\cite{cao2018openpose}, 3D poses~\cite{pavllo20193d}, and 3D motion reconstruction~\cite{Mehta:2017}.
Recently, Aberman~\etal~\shortcite{aberman2019learning} extracted character-agnostic motion, view angle, and skeleton, as three disentangled latent codes, directly from videos, bypassing 3D reconstruction. 
Existing methods that extract motion from videos~\cite{Mehta:2017, kanazawa2019learning} are typically not concerned with style, while our work is the first to extract the style, rather than the motion, from video-captured human motion samples, bypassing 3D reconstruction.

Another stream of work is focused on motion transfer in video using deep learning techniques. The existing approaches provide a different perspective on how to extract motion in 2D \cite{chan2019everybody,aberman2019deep} or 3D \cite{liu2018neural} from one video, and apply to it the appearance of a target actor from another video.

\section{Motion Style Transfer Framework}
\label{sec:FSMST}
Our framework aims at translating a motion clip of an animated character with a given \emph{content} to another motion that exhibits the same content, but performed using a different \emph{style}.
The desired \emph{target style} may be inferred from a few (or even a single) 3D motion clips, or a video example. Importantly, the target style might not be one of those seen during training.
We only assume that each of the motion clips in the training set is assigned a style label; there's no pairing requirement, i.e., no need for explicit pairs of motions that feature the same content performed using two different styles.

\begin{figure}
	\centering
	\includegraphics[width=\linewidth]{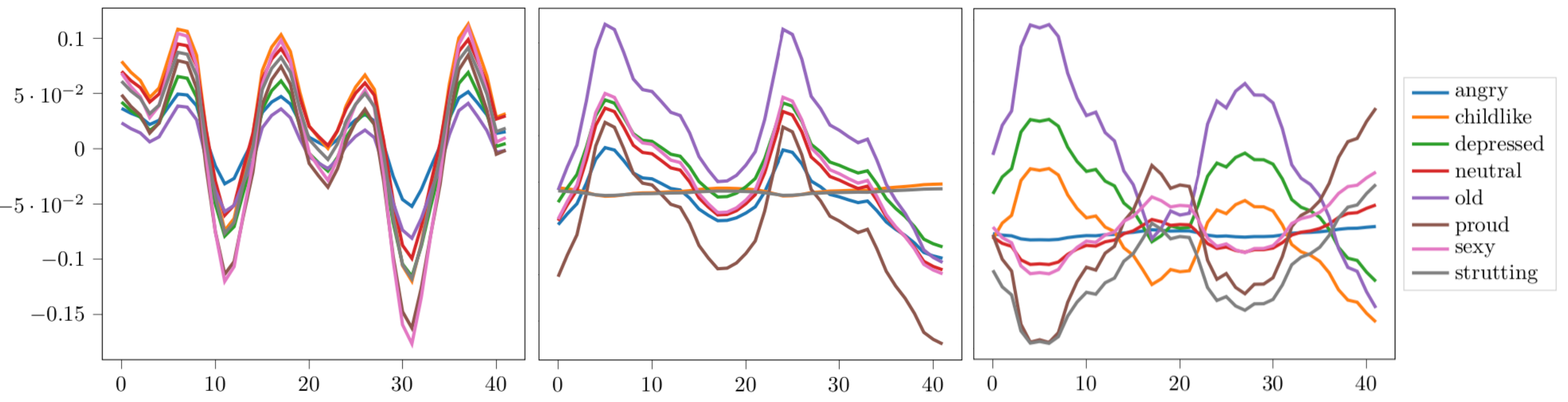} 
	\caption{Visualization of deep features in our decoder. Each plot visualizes a specific channel of a deep feature as eight different styles are applied to the same motion content. It can be seen that the signals differ only by a temporally-invariant affine transform and that the signal shape (per channel) is preserved. Thus, the motion content is preserved as well.}
	\label{fig:adain_deep_features}
\end{figure}

We treat style transfer as a conditional translation model, and propose a neural network that learns to decompose motion into two disentangled latent codes, a temporal latent code that encodes motion content, and a temporally-invariant latent code that encodes motion style.
These two codes affect the generated motions via two different mechanisms.
The content code is decoded into a motion by applying a sequence of temporal convolutional layers, each yielding a set of temporal signals that determine the joint rotations in a high-dimensional feature space.
The style code is used to modify only the means and variances of the generated temporal deep features via temporally-invariant adaptive instance normalization (AdaIN), thereby preserving the content of the original motion. Figure~\ref{fig:adain_deep_features} visualizes three channels of deep features generated by our network's decoder, showing the effect of different styles applied to the same motion content. 

\begin{figure*}
	\centering
	\includegraphics[width=\linewidth]{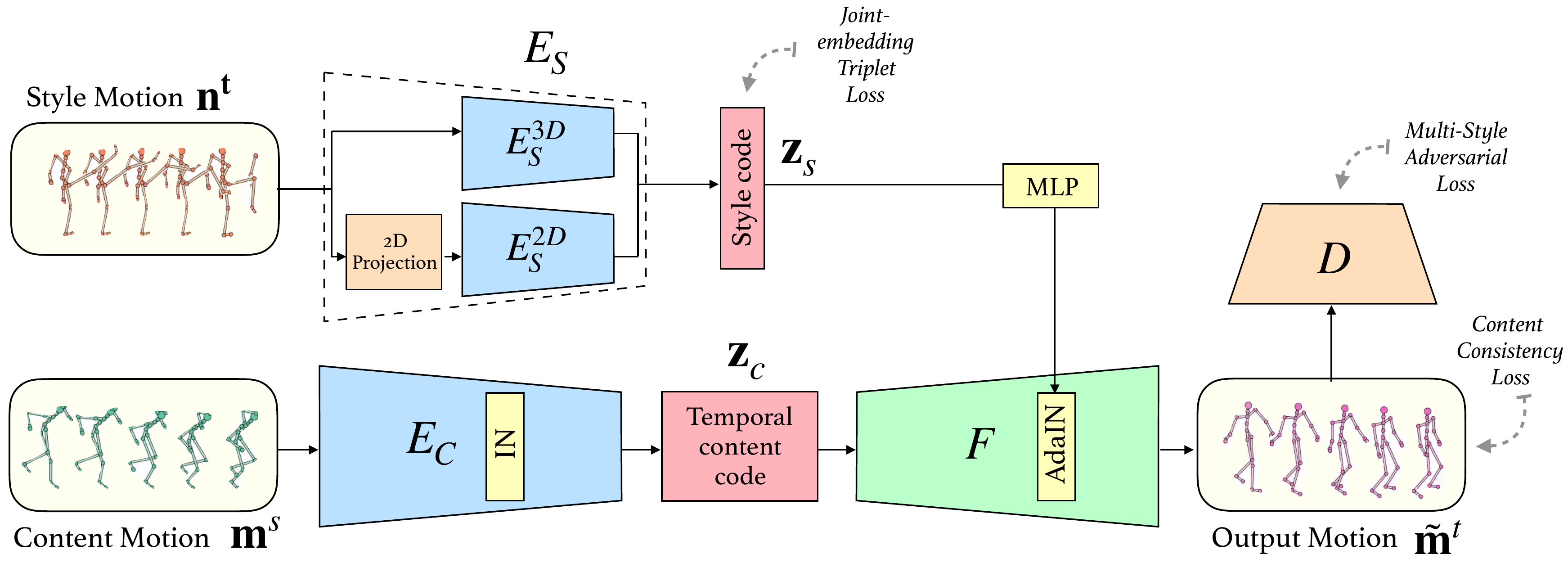} 
	\caption{Our motion-style transfer architecture consists of encoders for content ($E_C$) and style ($E_S$), which extract a latent content code ($\bz_c$) and a latent style code ($\bz_s$), respectively, where the style code can be extracted from either 3D motions (via $E_{S}^{3D}$) or 2D projections ($E_{S}^{2D}$).
		During encoding, the content code is stripped of style by instance normalization (IN) layers. The content code is then used to reconstruct a motion by a decoder $F$, which contains AdaIN layers that modify temporally-invariant second order statistics of the intermediate deep features decoded by $F$. The output motions are fed into a multi-style discriminator $D$ that judges whether it belongs to a certain style. 
	}
	\label{fig:MST_arc}
\end{figure*}

In addition, in order to support style extraction from videos, we learn a joint embedding space for style by extracting style codes from either 3D or 2D joint positions, using two different encoders, which are encouraged to map 3D motions and their 2D projections into the same code. 

Figure~\ref{fig:MST_arc} describes the high-level architecture of our framework, whose  various components are described in more detail below. 

\subsection{Architecture}
Our framework, depicted in Figure~\ref{fig:MST_arc}, consists of a conditional motion translator that takes as input two motion clips: the content motion $\bm^s$, with a source style $s\in \mms$, as well as a style motion $\bn^t$, with a target style $t\in \mms$.
The output motion $\tilde{\bm}^t$ is supposed to consist of the content of $\bm^s$, performed using style $t$.
We next describe the design and role of the different components in our framework.

\paragraph{Motion Representation}
In our setting, a motion clip $\bm\in\mathbb{R}^{T\times d}$ is a temporal sequence of $T$ poses where each pose is represented by $d$ channels. We choose to represent the content input $\bm^s$ and the style input $\bm^t$ differently from each other. Since a motion is well defined by the joint rotations (as opposed to joint positions), and since the content input is strongly correlated with the output motion, we represent $\bm^s$ using rotations (unit quaternions), i.e., $\bm^s \in \mathbb{R}^{T\times 4J}$, where $J$ is the number of joints.
In contrast, since style can be inferred from the relative motion of joint positions, and to facilitate learning a joint embedding space for 3D and 2D motions, we represent the style input using joint positions ($\bn^t \in \mathbb{R}^{T\times 3J}$). The output motion $\tilde{\bm}^t$ is represented using joint rotations, which enables the extraction of standard character animation files without any further post-processing. Note that the global root positions are discarded from the representation of the network's input/output, and are treated separately during test time, as explained later in this section.

\paragraph{Motion Translator}
The motion translator consists of a content encoder $E_C$, a style encoder $E_S$, and a decoder $F$, where $E_C$ and $E_S$ encode the input into two latent codes, $\bz_c$ and $\bz_s$, respectively.
$E_C$ consists of several temporal, 1D convolutional layers~\cite{holden2015learning}, followed by several residual blocks that map the content motion to a temporal content latent code $\bz_c$. During encoding, the intermediate temporal feature maps undergo instance normalization (IN), which effectively ensures that the resulting content code is ``stripped'' of style.

The encoding of style is performed by one of the two encoders $E_{S}^{2D}, E_{S}^{3D}$, depending on whether the joint coordinates are in 2D (extracted from a video) or in 3D.
We assume that the style does not change in mid-motion, and use a sequence of 1D convolutional layers to map $\bn^t$ into a fixed-size (independent on the temporal length of the clip) latent code $\bz_s$.
We'd like 2D and 3D motions performed with the same style to be mapped to the same latent vector $\bz_s$.
Thus, at each iteration during training we use a 3D motion clip, along with a 2D perspective projection of that same clip, and feed both clips into the corresponding encoders. The latent code $\bz_s$ is then obtained by averaging the output of the two encoders.
At test time, the style code is extracted by only one of the two encoders, depending on the type of style input.

The decoder $F$ consists of several residual blocks with adaptive instance normalization (AdaIN)~\cite{huang2017arbitrary}, followed by convolutional layers with stride that upsample the temporal resolution of the clip. The AdaIN layer constitutes a normalization layer that applies an affine transform to the feature activations (per channel). For each AdaIN layer with $c$ channels, the network learns a mapping (a multilayer perceptron, or MLP) of the style code $\bz_s$ into $2c$ parameters that modify the per-channel mean and variance.

Note that the affine transformation is temporally invariant and hence only affects non-temporal attributes of the motion. Thus,
while the content encoder effectively removes the source style $s$ by normalizing the non-temporal attributes with IN, the decoder injects the target style $t$ by using AdaIN to scale and shift the feature channels to target values inferred from the style code.

In summary, using the notation introduced above, our conditional motion translator $G$ may be formally expressed as:
\begin{equation}
\tilde{\bm}^t = G\left(\bm^s \vert \bn^t \right) = F \left( E_C(\bm^s) \vert E_S(\bn^t) \right).
\end{equation}

\paragraph{Multi-Style Discriminator}
Our discriminator $D$ follows the multi-class discriminator baseline proposed at~\cite{liu2019few}. $D$  is a single component that is trained to cope with $|\mms|$ adversarial tasks simultaneously, where each task aims to determine whether an input motion is a real motion of a specific style $i\in\mms$, or a fake output of $G$. When updating $D$ for a real motion of source style $i\in\mms$, $D$ is penalized if its $i$-th output is false.  For a translation output yielding a fake motion of source style $i$, $D$ is penalized if the $i$-th output is positive. Note that $D$ is not penalized for not predicting false for motions of other styles. When updating $G$, it is penalized only if the $i$-th output of $D$ is false. 

A detailed description of the architecture layers and parameters is given in the Appendix.

\paragraph{Global Velocity}
At test time, we aim to extract the desired style even from a single example. 
However, since the content of the two input motions (style and content) may be different, the translation of global velocity, a property which is often correlated with style, is a challenging task. For example, in order to convert neutral walking to old walking, the global velocity should decrease. However, inferring such a property form a single clip of old kicking, where the root position is nearly static, is practically impossible, especially when the style is previously unseen. A principled solution to this problem is outside the scope of this work; below, we describe a heuristic solution that has worked well in our experiments. In this solution, the root positions of our output motion are directly taken from the content input sequence. However, since global velocity is correlated with style, we perform a dynamic time warping on the global velocity, based on the velocity ratio between the two input motions. More precisely, for each motion sequence, we measure the velocity factor as the temporal average of the maximal local joint velocity,
\begin{equation}
V= \frac{1}{T}\sum_{\tau=1}^{T} \max_{j\in J}\{v^j(\tau)\},
\end{equation}
where $v^j(\tau)$ is the local velocity of the $j$-th joint in time $\tau$.
Next, we warp the temporal axis by the factor $V^{\text{sty}}/V^{\text{con}}$, where $V^{\text{sty}}$ and $V^{\text{con}}$ are the velocity factors for the style and content inputs, respectively. 
We find that in most of our examples local joint velocity captured style information better than global velocity, and for that reason decided to use the above definition for the velocity factor.

Note that this global transformation is reversible, such that if we use the output motion as the content input, and the original content motion as the style input, we recover the global velocity of the original motion.

\paragraph{Foot Contact}
As our network is built upon 1D temporal convolution layers, raw outputs tend to suffer from foot skating artifacts. In order to cope with the issue, we extract foot contact labels from the content input, use them to correct the feet positions and apply IK to fix the corresponding output poses (before the global velocity warping). As a result, we get visually plausible outputs, with no foot skating.
While this fix works well in most cases, note that it assumes that the foot contact timing is part of the content, and not of the style. However, this is not always true: consider, for example, a zombie walking while dragging one of the feet.
This aspect should be further addressed in future work.

\subsection{Training and Loss}
Our dataset is trimmed into short overlapped clips, which comprise our motion collection $\mm$. However, note that at test time the length of the input sequences can be arbitrary, since the networks are fully convolutional. 

Note that although the content input motion and the output motion are both represented using joint rotations, all of the losses are applied to joint positions as well.
This is done by applying a forward kinematics layer \cite{villegas2018neural,pavllo2019modeling} on the aforementioned components, which for simplicity, is not explicitly mentioned in the following equations.

\paragraph{Content Consistency Loss}
In case that the content input $\bm^s$ and the style input $\bn^t$ share the same style ($t=s$), it is expected that the translator network will constitute an identity map, regardless of the content of $\bn^t$.
Thus, in every iteration we randomly pick two motion sequences from our dataset $\mm$, with the same style label, and apply the content consistency loss which is given by
\begin{equation}
\Loss_{\text{con}} = \bbe_{\bm^s,\bn^s\sim\mm} \Vert  F\left(E_C(\bm^s) \vert E _S(\bn^s) \right) - \bm^s   \Vert_1,
\label{eq:content_loss}
\end{equation}
where $\Vert\cdot\Vert$ represents the $L_1$ norm. Note that when $\bn^s=\bm^s$, Equation \eqref{eq:content_loss} becomes a standard reconstruction loss.

\paragraph{Adversarial Loss}
Since our training is unpaired, the adversarial loss is the component which is responsible, in practice, to manipulate the style of $\bm^s$ via
\begin{eqnarray}
\Loss_{\text{adv}} &=& \bbe_{\bn^t\sim\mm} \Vert D^t(\bn^t) - 1  \Vert^2 \\ \nonumber
&+& \bbe_{\bm^s,\bn^t\sim\mm\times\mm}  \Vert D^t(F\left(E_C(\bm^s) \vert E _S(\bn^t) \right)  \Vert^2 ,
\label{eq:adv_loss}
\end{eqnarray}
where $D^t(\cdot)$ represents the discriminator output which corresponds to the style class $t\in\mms$.
In order to stabilize the training of the generator in the multi-class setting, we apply feature matching loss as regularization~\cite{wang2018high}. This loss minimizes the distance between the last feature of the discriminator when fed by a real input of a specific style (averaged over the set) to the same feature when fed by a fake output of the same target style, via 
\begin{equation}
\Loss_{\text{reg}} = \bbe_{\bm^s,\bn^t\sim\mm\times\mm}  \Vert D_f(\tilde{\bm}^t) - \frac{1}{|\mm_t|}\sum_{i\in\mm_t}D_f(\bn^i_t) \Vert_1,
\label{eq:reg_loss}
\end{equation}
where $\mm_t$ is a subset of motions with style $t\in\mms$ and $D_f$ is a sub-network of $D$ that doesn't include the prediction (last) layer.

\subsection{Joint 2D-3D Style Embedding}
Intuitively, the style of motion can be identified by observing the character 3D positions in space-time as well as their 2D projections using a view with a reasonable elevation angle.
While handcrafted representations are designed to treat a single form of input, we exploit the power of deep learning to learn a joint embedding space for extracting style from both 3D and 2D motion representations. The learned common latent space can be used for various tasks, such as video-based motion style retrieval, style interpolation and more. In our context, we exploit the properties of this space to extract style directly from real videos during test time, while bypassing the need for 3D reconstruction, an error-prone process which adds noise into the pipeline.

\paragraph{Joint Embedding Loss}
In order to construct a common latent space for style, our loss encourages pairs of 3D-2D motions to be mapped into the same feature vector by
\begin{equation}
\label{eq:joint_embedding}
\Loss_{\text{joint}}  =  \bbe_{\bn^t\sim \mm}\Vert E_{S}^{3D}(\bn^t) - E_{S}^{2D}(P(\bn^t; p))\Vert^2,
\end{equation}
where $P$ is a weak perspective projection operator that projects the input to a camera plane, with camera parameters $p$, that consist of scale $s$ and the Euler angles $v=( v^{\text{pitch}}, v^{\text{yaw}}, v^{\text{roll}})$. For each motion clip, we define the local Z-axis to be the temporal average of per-frame forward directions, which are computed based on the cross product of the Y-axis and the average of vectors across the shoulders and the hips. During training  $v^{\text{roll}} = v^{\text{pitch}} = 0$ are fixed while $v^{\text{yaw}} \in [-90^\circ, 90^\circ]$ and $s \in [0.8,1.2]$ are randomly sampled five times in each iteration, to create five projections that are transferred to $E_{S}^{2D}$.

\paragraph{Style Triplet Loss}
In order to improve the clustering of the different styles in the latent space, we exploit the style labels and use the technique suggested by Aristidou \etal~\shortcite{aristidou2018deep} to explicitly encourage inputs with similar style to be mapped tightly together, by applying a triplet loss on the style latent space via
\begin{align}
\label{eq:triplet}
\Loss_{\text{trip}} & = & \bbe_{\bn^t, \bx^t,\bw^s \sim \mm}
[ &\Vert E_S(\bn^t) - E_S(\bx^t)\Vert  \,- \\ \nonumber
& & &\Vert E_S(\bn^t)- E_S(\bw^s)\Vert \,+\, \delta ]_{+}, 
\end{align}
where $\bx^t$ and $\bw^s$ are two motions with different styles $s \neq t$, and $\delta=5$ is our margin. This loss encourages the distance between feature vectors of two motion inputs that share the same style to be smaller, at least by $\alpha$, than the distance between two motions with different styles.

Our final loss is given by a combination of the aforementioned loss terms:
\begin{equation}
\label{eq:loss}
\Loss  =  \Loss_{\text{con}}  +  \alpha_{\text{adv}}\Loss_{\text{adv}}  +   \alpha_{\text{reg}}\Loss_{\text{reg}}   +  \alpha_{\text{joint}}\Loss_{\text{joint}}  +  \alpha_{\text{trip}}\Loss_{\text{trip}},
\end{equation}
where in our experiments we use $\alpha_{\text{adv}}=1$,  $\alpha_{\text{reg}}=0.5$, $\alpha_{\text{joint}}=0.3$ and $\alpha_{\text{trip}}=0.3$.
\section{Discussion}

As discussed in Section~\ref{sec:rel}, adjusting the mean and variance of deep feature channels in a neural network proved to be effective for manipulating the style of 2D images. Although motion style is a conceptually and visually different notion, we have shown that a similar mechanism can be used to manipulate the style of motion sequences.

Below we show that our technique may be seen as a generalization of the style transfer technique of Yumer and Mitra~\shortcite{yumer2016spectral}. Their technique is based on pairs, while ours is unpaired, and uses learning to deal with unseen styles. However, we present a derivation that shows the commonality in the building blocks, the analysis of the motion, and how style is transferred. These commonalities contribute to the understanding of our method.

In their work, Yumer and Mitra~\shortcite{yumer2016spectral} propose a method for motion style transfer in the frequency domain. They show that given two motions $\by^s$ and $\by^t$ with similar content and different styles $s \neq t$, a new, arbitrary, motion $\bx^s$ with style $s$ can be transferred to style $t$. The transfer result is given in the frequency domain by
\begin{equation}
\label{eq:yumer_freq}
\tilde{\bx}^t(\omega) = \vert \tilde{\bx}^t(\omega) \vert e^{i\measuredangle \tilde{\bx}^t(\omega)},
\end{equation}
where the magnitude of the output is given by
\begin{equation}
\label{eq:yumer_mag}
\vert \tilde{\bx}^t(\omega) \vert \;= \vert \bx^s(\omega)\vert +   \vert \by^t(\omega)\vert -  \vert \by^s(\omega)\vert  \nonumber
\end{equation}
and the phase function is taken from the original input signal $\measuredangle \tilde{\bx}^t(\omega) = \measuredangle\bx^s(\omega)$. Applying the inverse Fourier transform to Equation \eqref{eq:yumer_freq}, we get
\begin{equation}
\label{eq:yumer_conv}
\tilde{\bx}^t(\tau)\approx \bx^s(\tau) + g^{s\rightarrow t}(\tau) \ast \bx^s(\tau), 
\end{equation}
where $g^{s\rightarrow t}(\tau)$ is a convolution kernel (in time, $\tau$) whose weights depend on the source and target styles. 

Equation \eqref{eq:yumer_conv} implies that the approach of Yumer and Mitra \shortcite{yumer2016spectral} may be implemented using a convolutional residual block to modify the style of a motion sequence, where the target style is defined by the weights of the kernel $g^{s\rightarrow t}$, which depend on the source style $s$ and well as the target style $t$.

Similarly, our style translation framework also uses convolutional residual blocks as its core units, where the kernel weights effectively depend on the source and target style. Although the kernels are fixed at test time, their weights are effectively modified by the IN and AdaIN layers, as a function of the styles $s$ and $t$. 
Convolution in time (with kernel $k(\tau)$ and bias $b$) followed by an IN/AdaIN layer can be expressed as:
\begin{equation}
\label{eq:conv_IN}
\tilde{\bx}(\tau) = \beta \left[\bx(\tau) \ast k(\tau) + b\right] + \gamma = \bx(\tau) \ast \beta k(\tau) + \beta b + \gamma,
\end{equation}
where $\beta$ and $\gamma$, are the IN or AdaIN parameters.
Equation \eqref{eq:yumer_conv} is, in fact, equivalent to a single convolution with an effective kernel weights $\hat{k}(\tau) = \beta k(\tau)$ and bias $\hat{b} = \beta b(\tau) + \gamma$, that are modified as a function of some input. For the IN case, $\beta$ and $\gamma$ depend on the input signal $\bx^s$, since they are calculated such that the output has zero mean and unit variance. In the AdaIN case, the parameters are produced by an MLP layer as a mapping of the target style's latent code. Thus, the use of IN and AdaIN in our architecture effectively controls the convolutions by the source and target styles.

Rather than directly transferring the motion style from $s$ to $t$ with a single convolution, as in Equation \eqref{eq:yumer_conv}, our process may be viewed as consisting of two steps: first, the source style is removed by the encoder $E_C$, which depends only on $s$, and then the target style is applied by the decoder $F$, via the AdaIN layers, which depend only on $t$.
Thus, our network performs style transfer in a modular way, making it easier to accommodate new styles. In contrast, using a single kernel makes it necessary to learn how to translate each style in $\mms$ to a every other style (all pairs).

\begin{figure*}
	\centering
	\includegraphics[width=\linewidth]{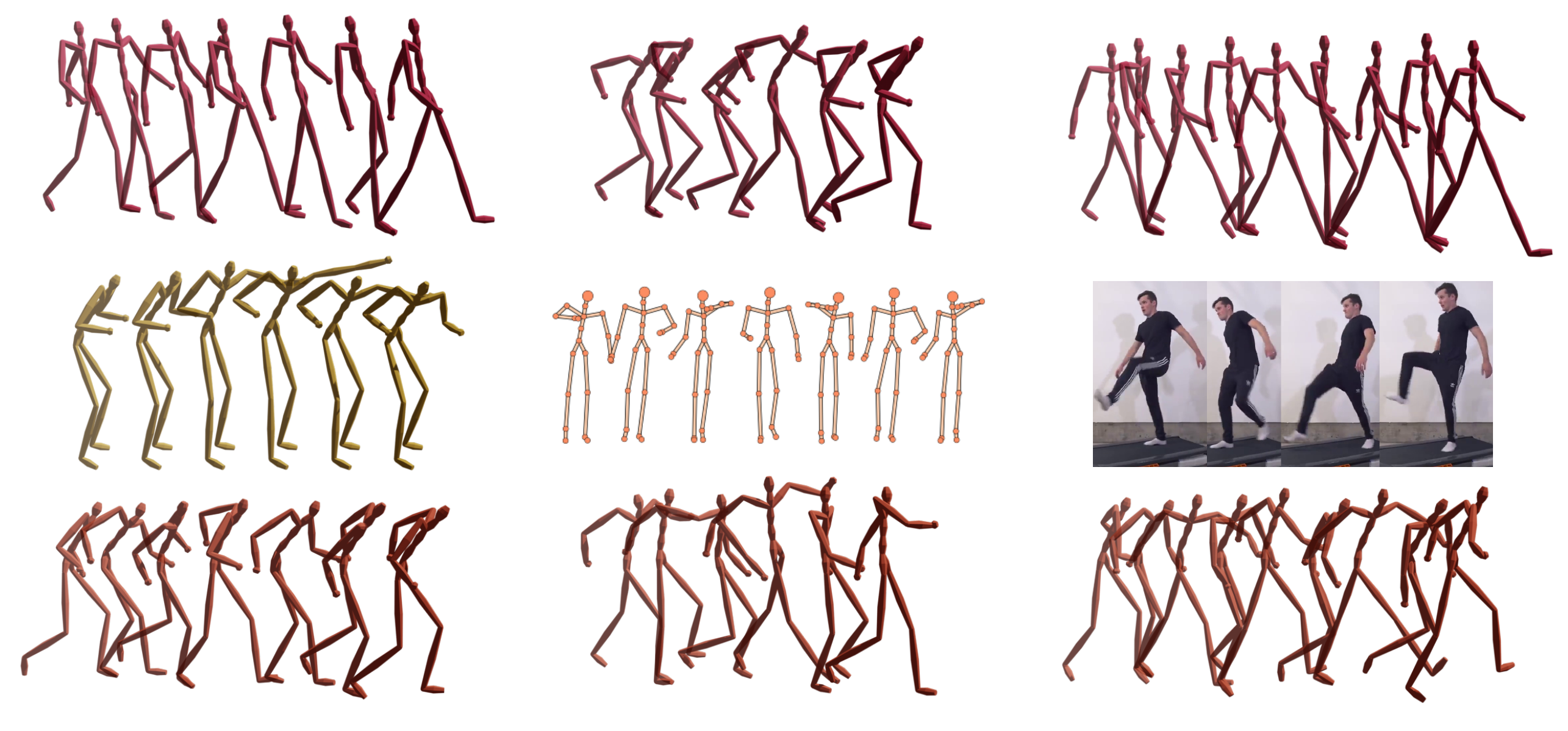} 
	\begin{tabular}{cc}
		(a) \hspace{5.7cm} (b)  \hspace{5.7cm} (c) 
	\end{tabular}
	\caption{Samples of our style transfer results. The motion of the content input (top row) is transferred to a motion with similar content and different style (bottom row), while the style can be extracted from various sources (middle row) such as 3D animated characters (a) 2D projection of 3D motions (b) and video sequences (c).}
	\label{fig:gallery}
\end{figure*}

\section{Experiments and Evaluation}
\label{sec:exp}
In this section we evaluate our method and perform various experiments and comparisons that demonstrate some interesting insights and interpretation of our style transfer mechanism.

Firstly, several samples of our style transfer results are shown in Figure~\ref{fig:gallery}. The full motion clips are included in the accompanying video. Note that our network outputs joint rotations, hence, our results do not require any further processing such as IK, and can be directly converted to the commonly used motion representation files, and visualized.
Although our joint rotations are represented by unit quaternions, which may lead to discontinuities within neural networks \cite{zhou2019continuity}, our output quaternions tend to be smooth due to the temporal 1D convolutions performed by our network.
Our results demonstrate that our system can transfer styles that are extracted from various sources, such as 3D animated characters, 2D projection of 3D motions and real videos, within a unified framework.
The examples demonstrating style transfer from a video use only a short (3-second) video clip as the sole (and previously unseen) style example. We are not aware of any other style transfer method with this capability.

\paragraph{Implementation Details}
We used two different datasets to perform our experiments. The first dataset,
supplied by Xia~\etal~\shortcite{xia2015realtime}, contains motion sequences that are labeled with eight style labels. The second, is our own newly captured dataset, which contains various motions performed by a single character in 16 distinct styles. For convenience, we refer to the datasets as A and B, respectively.
The motion sequences within each of the datasets are trimmed into short overlapping clips of $T=32$ frames with overlap of $T/4$, resulting in about 1500 motion sequences for dataset A and 10500 for B. In addition, the motions in each dataset are split into two disjoint train and test sets, with the test set consisting of 10\% of the samples.  

Our framework is implemented in PyTorch and optimized by the Adam optimizer. A training session takes about about 8 hours for dataset A, and double the time for dataset B, using an NVIDIA GeForce GTX Titan Xp GPU (12 GB).


\subsection{Latent Space Visualization}
In this experiment we project the content codes and style parameters of some random motion samples from dataset A onto a 2D space by using t-distributed stochastic neighbor embedding (t-SNE), and plot the results in order to gain a better understanding of how the network interprets content and style in practice.

\begin{figure}
	\centering
	\includegraphics[width=\linewidth]{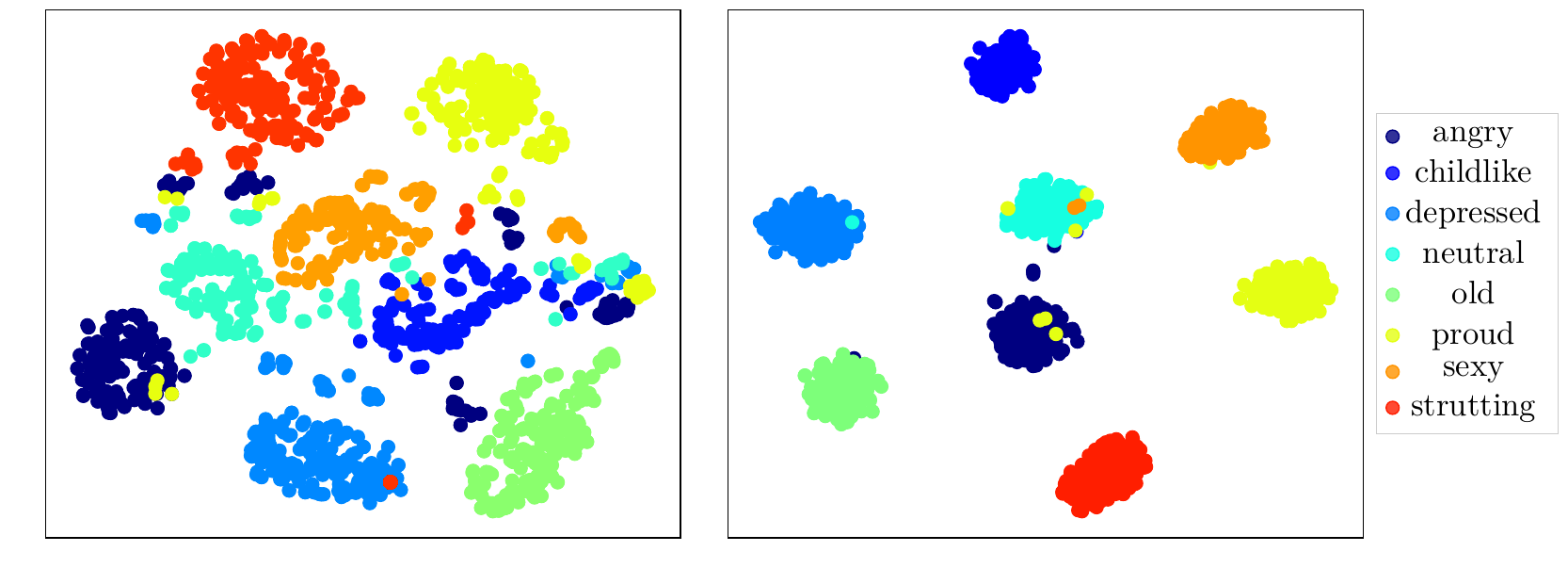} 
	\begin{tabular}{cc}
		(a) \hspace{3cm}&\hspace{3.3cm} (b) 
	\end{tabular}
	\caption{The AdaIN parameters extracted from the style codes are projected onto 2D space using t-SNE and colored based on their style labels. The system is trained without triplet loss (a) and with triplet loss (b). It can be seen that our framework learns to cluster the AdaIN parameters as a function of style label in both cases, while the addition of the triplet loss results in tighter clusters. }
	\label{fig:PCA_AdaIN_code}
\end{figure}

\begin{figure}
	\centering
	\includegraphics[width=0.6\linewidth]{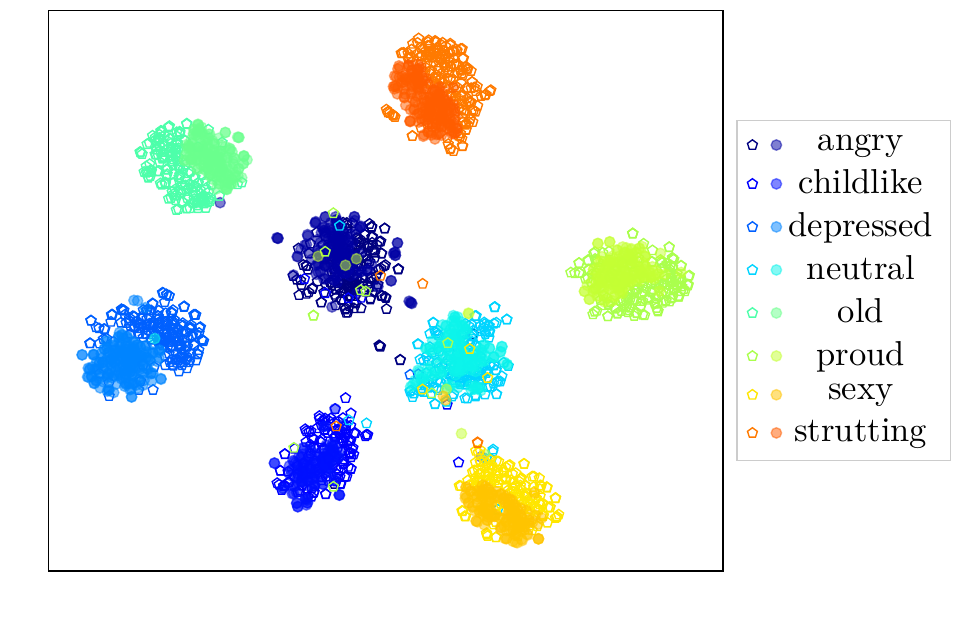} 
	\caption{Joint embedding of style codes parameters extracted from 3D motions as well as directly from 2D videos. 
		\label{fig:2D_3D_space}
	}
\end{figure}

\paragraph{Style Code}
Figure~\ref{fig:PCA_AdaIN_code} shows the 2D projection of our style parameters (AdaIN), where each sample is marked with a color corresponding to its style label. It can be seen that our network learns to cluster the style parameters, which means that style inputs that share the same style will manipulate the motion content in a similar way. This result demonstrates that the extracted style parameters mostly depend on the style label.
 
As previously discussed, our framework treats style as a set of properties, shared by motions in a group, which can be manipulated (added/removed) by an affine, temporally invariant, transformation (AdaIN) applied to deep features.
When such common properties exist within the group, the clusters are naturally formed even without the need for triplet loss (Figure~\ref{fig:PCA_AdaIN_code}(a)). However, since a given style may be described by different nuances for different content motions (e.g., proud boxing has some hand gestures that do not exist in proud walking), a triplet loss encourages (but does not enforce) style codes of the same group to be closer to each other. This loss emphasizes commonalities within the group, making the clusters tighter, as can be observed in Figure~\ref{fig:PCA_AdaIN_code}(b), and leads to better content-style disentanglement. 

Figure~\ref{fig:2D_3D_space} visualizes style codes parameters (AdaIN) extracted from 3D motions together with ones extracted from video. It may be seen that the latter codes, for the most part fall into the same clusters as the former ones.

\begin{figure}
	\centering
	\includegraphics[width=\linewidth]{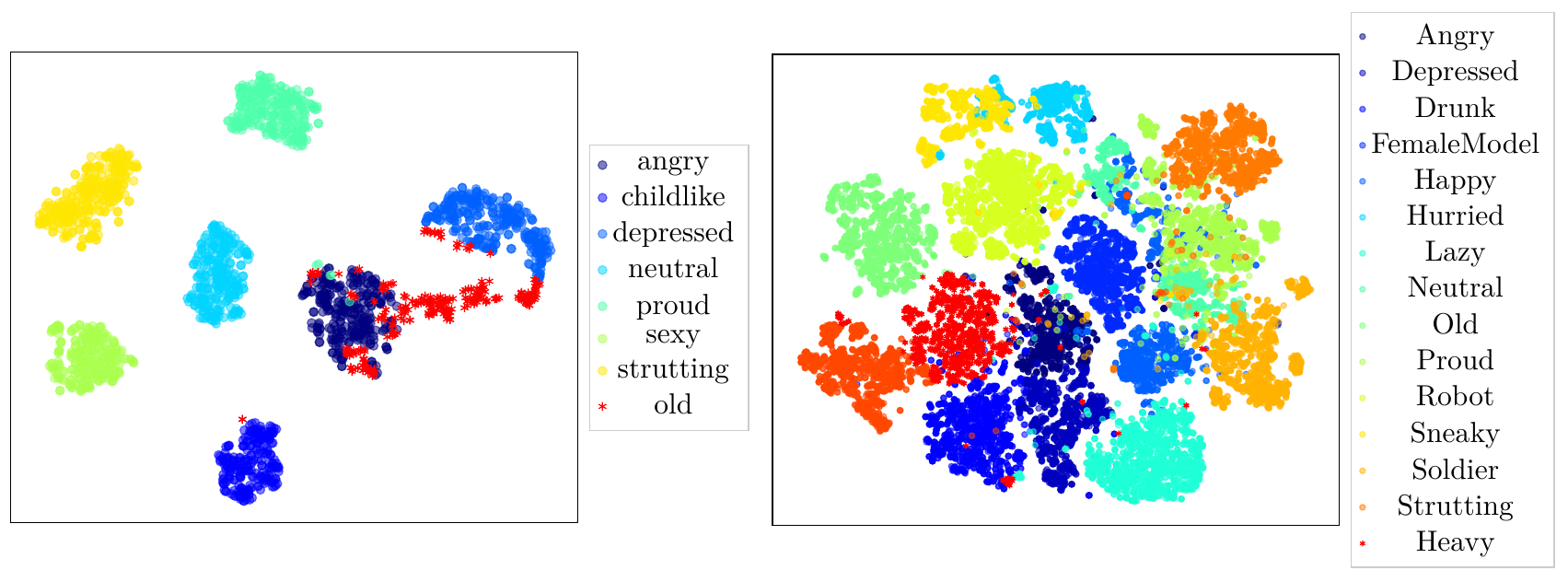} 
	\begin{tabular}{cc}
		(a) \hspace{3cm}&\hspace{3.6cm} (b) 
	\end{tabular}
	\caption{Unseen styles. (a) Trained on dataset A excluding the ``old'' style. (b) Trained on dataset B excluding the ``heavy'' style. It can be seen that a larger number of style classes enables the network to better generalize styles in the latent space}
	\label{fig:unseen_clusters}
\end{figure}

\paragraph{Unseen Styles}
Generally speaking, our network enables to extract styles from arbitrary motion clips during test time. However, in practice, when the number of seen styles is small, the network may overfit to the existing styles, as one might suspect when observing the well-separated clusters in Figure~\ref{fig:PCA_AdaIN_code}. We retrained our model with dataset A, excluding the motions that are labeled by the ``old'' style label, and then tested it using the motions within this group. Although the network successfully clusters the samples (Figure~\ref{fig:unseen_clusters}(a)), our results show that the style properties of the output are adapted from visually similar styles among those that were seen during training. For example, the ``old walking'' style code is close that that of ``depressed walking'', and the style transfer result indeed resembles that of ``depressed walking''.

We performed the same experiment with dataset B (which includes 16 styles), excluding the ``heavy'' style, and then tested the trained system with these motions.  As can be seen in Figure~\ref{fig:unseen_clusters} (b), the network again learns to cluster the test samples by their style labels. However, in this case, the output motions successfully adapted style properties from the new unseen clip, which means that the overfitting is significantly reduced when training on dataset B. This demonstrates that for the same framework with fixed dimensions, style can be generalized better, when there are more style classes, and that the network learns to identify properties that may be related to style, and to extract them even from unseen style inputs. However, when the number of styles is small, or the characteristic style properties are very different from those encountered during training, the network fails to generalize style.

The output motion clips of two experiments with unseen settings are shown in our supplemental video, next to an output that demonstrates how the same pair of inputs in each experiment is translated once the network has seen that style during training. As can be seen, although the outputs are different, in both cases the motion is plausible and the target style can be identified.

\begin{figure}
	\centering
	\includegraphics[width=\linewidth]{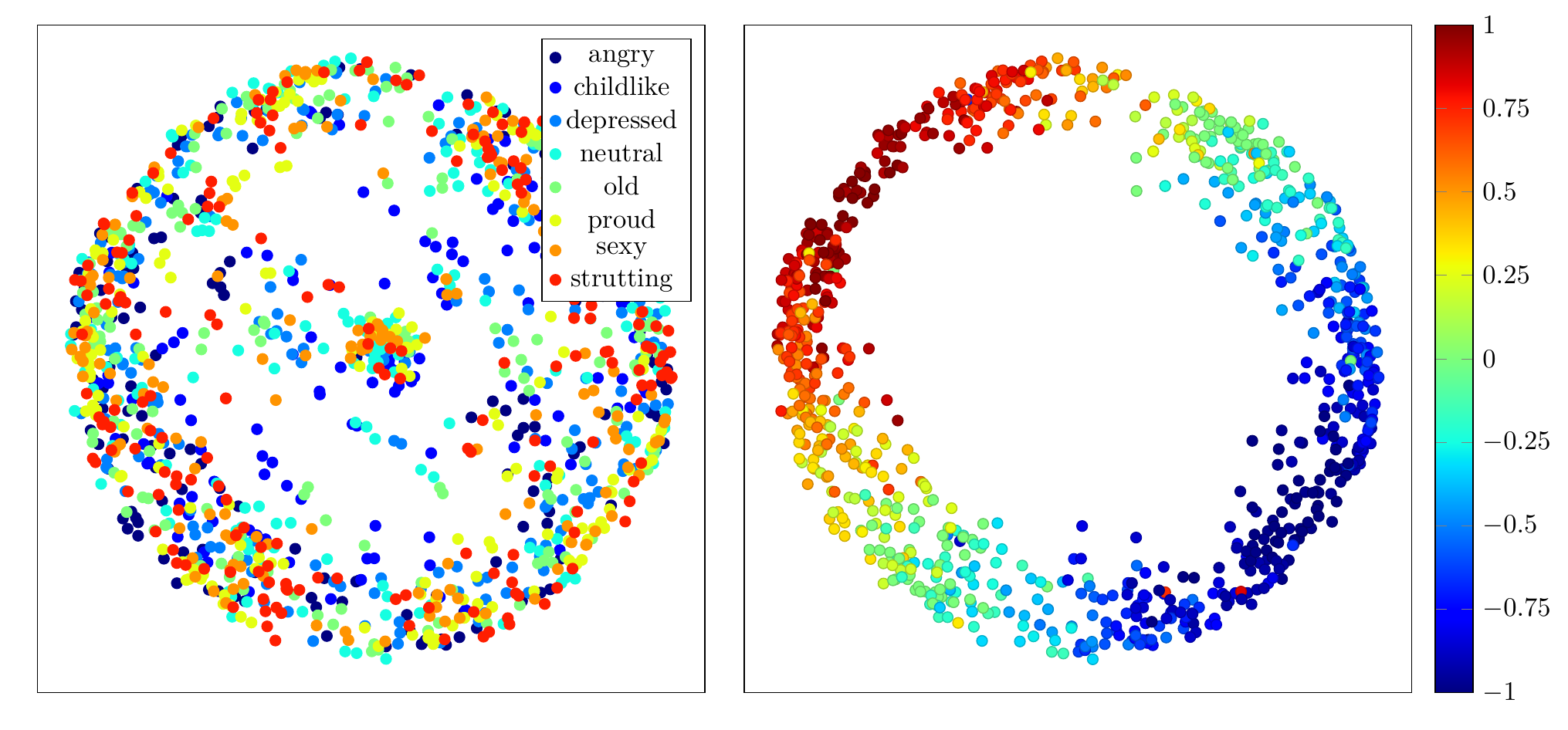} 
		\begin{tabular}{cc}
		(a) \hspace{3cm}&\hspace{3.6cm} (b) 
	\end{tabular}
	\caption{The content codes of our test samples are projected onto 2D space using PCA. (a) The samples are labeled by the style label. No clustering based on style label may be observed, suggesting that style information has been removed. (b) When visualizing only walking motions, while labeling samples by the phase of walking, it may be seen that our content code effectively parameterizes the motions using a single parameter -- the phase.}
	\label{fig:PCA_content_code}
\end{figure}

\paragraph{Content Code}
Figure~\ref{fig:PCA_content_code}(a) visualizes 2D projections of our content codes (dataset A), colored by their style labels. It can be seen that there is no clear correlation between the spatial positions of the points and their labels, which suggests that the content code probably does not contain significant style information.

Surprisingly, by observing the spatial distribution of the 2D points it can be seen that a subset of the samples forms a circle. The circle becomes nearly perfect by filtering out all the non-walking motions (using content motion labels that exist in the original dataset) and scaling the 2D space (the unscaled projection is elliptical).
For walking motion samples, the reduced space achieved by PCA captures 97.4\% of the original variation, which means that our projected content code preserves the information well. 

The nearly perfect circle is achieved due to 3 main reasons: (i) Walking motions in dataset A exhibit approximately the same velocity. (ii) The network discards global velocity and orientation (iii) Our motion samples are represented by a fixed size temporal window. Thus, the content of these periodic motions can be parameterized with a single parameter: the phase. In order to confirm this interpretation we calculate the period of each walking motion sample, extract the phase $\Theta$ of the middle frame, and color the corresponding point with $\sin(\Theta)$ in Figure~\ref{fig:PCA_content_code}(b). The continuous variation of the color along the circle suggests that our network effectively strips the style and represents walking motions with a single phase parameter. 

The phase representation for locomotion is well-known in character animation and is used, for example, to dictate the state or mode of a phase-based neural network that generates animations of humans \cite{holden2017phase}. 

\subsection{Comparison}
In this section we compare our approach to the method of Holden~\etal~\shortcite{holden2016deep} that performs style transfer by optimizing a motion sequence to satisfy two constraints, one for motion and one for style. Similarly to the seminal work of Gatys~\etal~\shortcite{gatys2016image} for image style transfer, the content is described by a set of deep features, and the style is represented by the Gram matrix of those features. However, while in the image domain the features are extracted by a classification network, here they are extracted by a motion autoencoder.

In order to perform the comparison the approaches are qualitatively evaluated by a user study that measures a few aspects of style transfer approaches. The results are evaluated with styles extracted from 3D motions, as well as from videos. However, since Holden~\etal~\shortcite{holden2016deep} extract styles only from 3D motions, we use a state-of-the-art 3D pose estimation algorithm \cite{pavllo20193d} to recover 3D poses, when the provided style input is a video. For a fair comparison we use dataset A, which is part of the CMU dataset \cite{CMU:mocap}, which Holden~\etal~\shortcite{holden2016deep} used to train their model.

\begin{figure*}
	\centering
	\includegraphics[width=\linewidth]{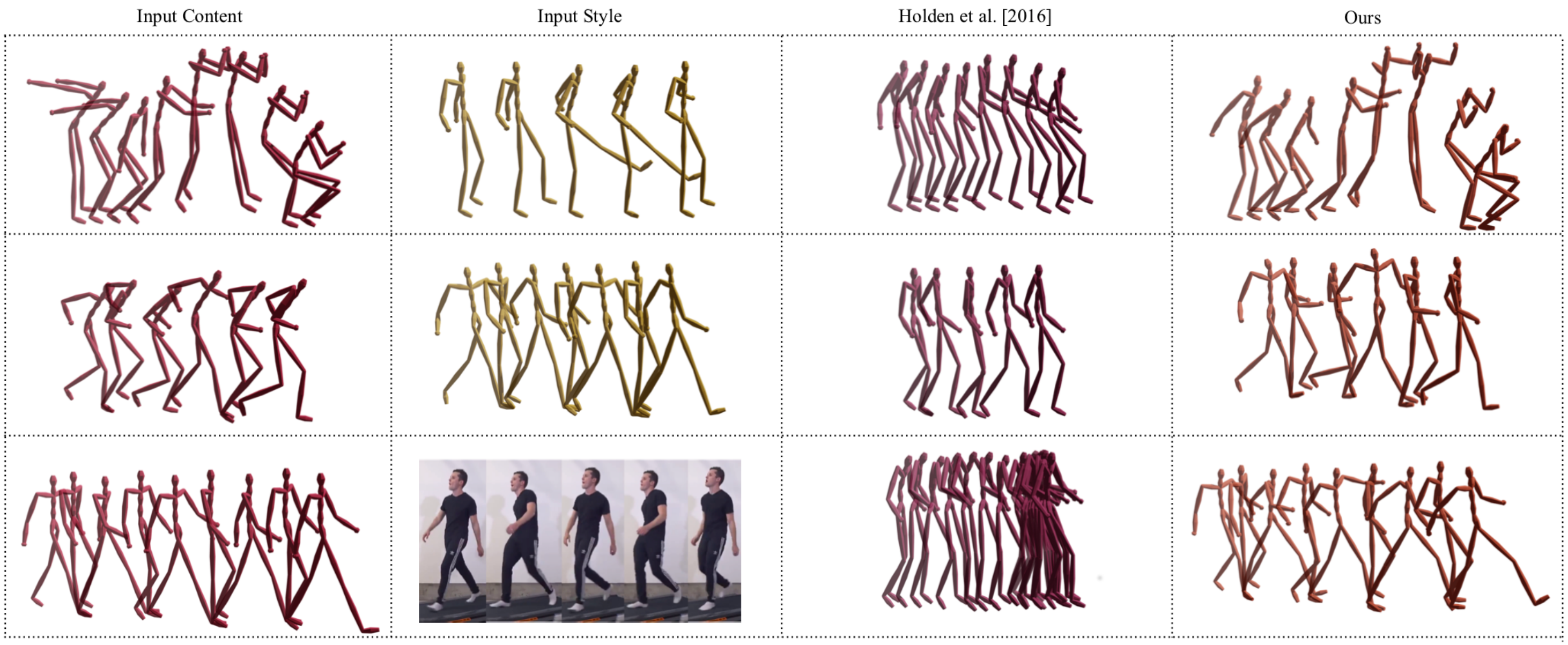} 
	\caption{Qualitative comparison of our method to the approach of Holden~\etal~\shortcite{holden2016deep}. The content input is shared across all the examples (each column shows a different example), 
	the input style is depicted in the first row, while the results of Holden~\etal~\shortcite{holden2016deep} and ours are given in the second and last row, respectively. We picked a fixed set of key frames of each motion to demonstrate the results. The full video sequences and more results can be found in the supplemental video.
	}
	\label{fig:qual_comparison}
\end{figure*}

A few results extracted from the full comparison given in our supplementary video are depicted in Figure~\ref{fig:qual_comparison}.

\subsubsection{User Study}
We performed a user study to perceptually evaluate the realism, style expressiveness and content preservation of our transfer results, while the style is extracted both from 3D motions and videos. 22 subjects were asked to answer a questionnaire with three types of questions, which we describe below. 

\paragraph{Realism}
In this part, we evaluate the realism of different motions. Users were presented with a pair of motions, both depicting the same type of content and style (e.g., angry jump). The motions were taken from three different sources: (1) Our original MoCap dataset, (2) Results of Holden~\etal~\shortcite{holden2016deep}
(3) Our results. Note that (2) and (3) are generated with similar inputs. 
Users were asked questions of the form: ``Which of the above motions look like a more realistic old walk?'', and had to choose one of the four answers: Left, Right, Both, or None.

132 responses were collected for this question type. 
Table \ref{tab:user_study_real} reports the realism ratios for each motion source. It may be seen that 75\% of our results were judged as realistic, which is a significantly higher ratio than the one measured for Holden \etal~\shortcite{holden2016deep}, and not far below the realism ratio of real MoCap motions.

\begin{table}[]
    \centering
    \begin{tabular}{ccc}
    \toprule
         MoCap &  Holden \etal~\shortcite{holden2016deep} & Ours \\
          \hline 
    79.17\% & 12.5\% & 75\% \\
    \bottomrule
    \end{tabular}\\ (a) \\
     
        \begin{tabular}{lcc}
           \toprule
          & Holden \etal~\shortcite{holden2016deep} & Ours \\
         \hline
      Content Preservation - 3D & 38.89\% & 61.11\% \\
      Content Preservation - video & 25\% & 75\% \\
      Style Transfer - 3D & 5.56\% &  94.44\% \\
      Style Transfer - video & 8.33\% & 91.67\%  \\
    \bottomrule
    \end{tabular}\\ (b) \\
    \caption{User study results. (a) Realism ratios. (b) Content preservation and style transfer ratings (\cite{holden2016deep} vs. Ours). The style inputs were either from 3D motion, or from video.}
    \label{tab:user_study_real}
    \label{tab:user_study_comp}
\end{table}


\paragraph{Content Preservation and Style Transfer}
In this part, we compare our style transfer results to those of Holden~\etal~\shortcite{holden2016deep} in terms of two aspects: the preservation of content and the transfer of style. Users were presented with a content input, a style input, and two transferred results, one by Holden \etal~\shortcite{holden2016deep} and the other by our method. They were asked to first select the motion whose content is closer to the content input (``Which of the motions on the right is more similar to the motion on the left in content?''), and then select the motion whose style is closer to the style input (``Which of the motions on the right is more similar to the motion on the left in style?'').  

110 responses were collected for each of these two questions. The results are reported in Table \ref{tab:user_study_comp}. 
The results indicate that our method was judged far more successful in both aspects (content preservation and style transfer), both when using a 3D motion as the style input, and when using a video. The reasons to which we attribute the large gaps in the ratings are discussed in detail later in this section.

\begin{figure}
	\centering
	\includegraphics[width=\linewidth]{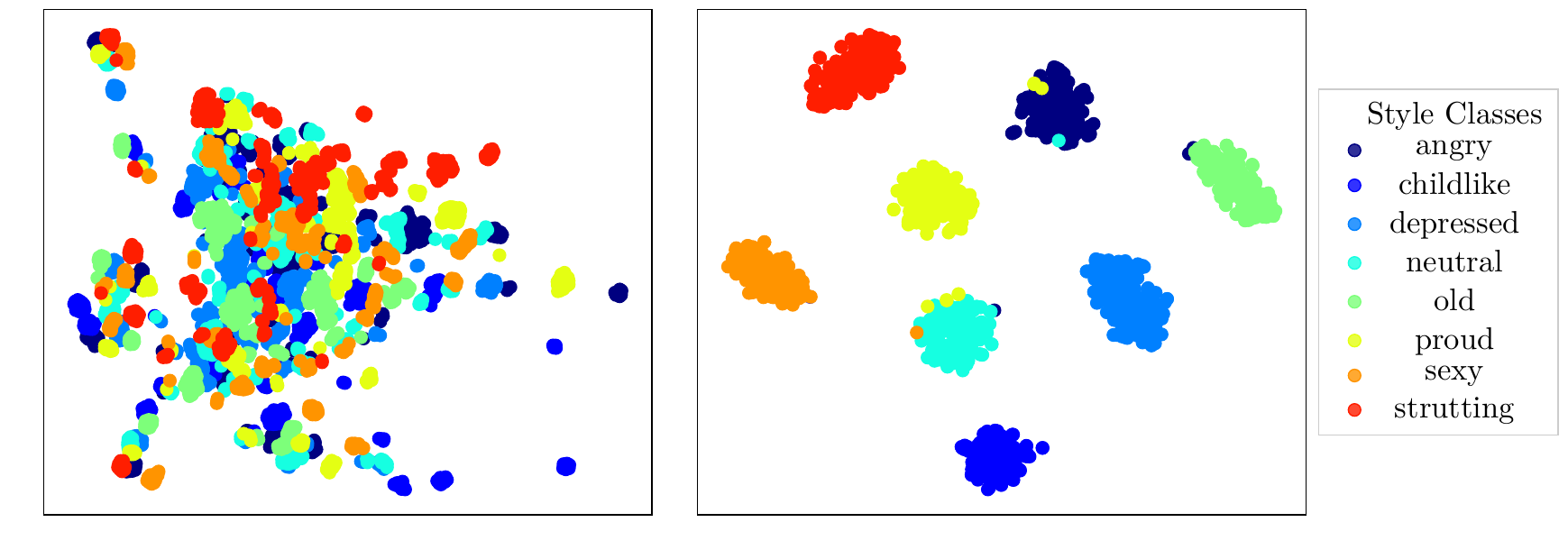} 
	\caption{Style codes extracted by our method (right) compared to the style representation of Holden~\etal~\shortcite{holden2016deep} (left). While our style codes are clustered by style labels, while in Holden's representation the style representation for many small scattered clusters, which implies dependencies between style and content. While our style codes are clustered by style labels, the style representation of Holden~\etal~\shortcite{holden2016deep} forms many small scattered clusters, which imply dependencies between style and content.
	}
	\label{fig:style_code_vs_others}
\end{figure}

It can be seen that the user study states that our method yields results which are more faithful to the task of style transfer. In particular, it can be seen that the approach of Holden \etal~\shortcite{holden2016deep} struggles to transfer style when the content of the two input motions is different (for example, when the input content motion is ``proud walking" and the input style is ``depressed kicking"). The main reason is that both content and style representations are derived from the same deep features, which leads to a dependency of content and style. In order to get a better understanding of their style representation, we projected the styles extracted by both methods into 2D, using PCA. Figure~\ref{fig:style_code_vs_others} shows the resulting maps. It can be seen that while our samples are clustered by the style labels (right plot), this cannot be observed for Holden's representation, which results in a multitude of small clusters, scattered over the 2D plane. Thus, while Gram matrices of features extracted by an autoencoder enable some degree of style transfer, they are clearly affected by other information present in the motion samples.

Moreover, we use video style examples, where a person demonstrates different walking styles, while walking on a treadmill. When poses are extracted from such a video, the root velocity is very small. In contrast, most of the content inputs have significant root velocity. This discrepancy poses no problem for our approach, but it adversely affects the method of Holden~\etal\shortcite{holden2016deep}, which is limited to work with input pairs that share the same content. 



Our method explicitly attempts to extract a latent style code from an input style motion, which enables clustering of motions with different content but similar style in the latent style space, thereby disentangling style from content. In contrast, Holden \etal~\shortcite{holden2016deep} represent style using Gram matrices, similarly to the seminal work of Gatys~\etal~\shortcite{gatys2016image} for image domain style transfer. In order to demonstrate the difference between the resulting style representations, we project the styles extracted by both methods into 2D, using PCA, and the results are shown in Figure \ref{fig:style_code_vs_others}.




\subsection{Ablation Study and Insights}
\paragraph{Effect of Adversarial Loss}
In this experiment we discarded the adversarial loss $\Loss_{\text{adv}}$ from our training. Surprisingly, our experiments show that the discriminator does not play the key role in the transferring of style.  
Furthermore, a single content consistency loss is sufficient to train the network to extract shared property with from labeled styles, and to cluster the style code samples by their style labels. However, we found that without the attendance of the adversarial loss, the perceived realism of the output motions is degraded, and artifacts such as some shaking can be observed. The comparison can be found in the supplementary video.

\paragraph{Style Code and Neutral Style}
In order to gain a better understating of the impact of the style code and the structure of its space, we neutralized the style branch by setting the AdaIN output to identity parameters (zero mean, unit variance). With these settings, the network outputs pure noise. The reason is that the network is trained in an end-to-end fashion, the scale and translation are also responsible for modifying the features such that the network outputs valid motions.
In addition, in order to understand whether the neutral style is more centralized in the style latent space than other styles, for every style label, we calculated the mean distance between its average style code to all the other average style codes.  We found that in both of the datasets the neutral style is among the top three styles in terms of that mean distance, which might suggest that the network learns to branch from neutral style into the other styles. However, we are not able to reach a definitive conclusion based on this experiment.

%




\subsection{Style Interpolation}

\begin{figure}
	\centering
	\includegraphics[width=\linewidth]{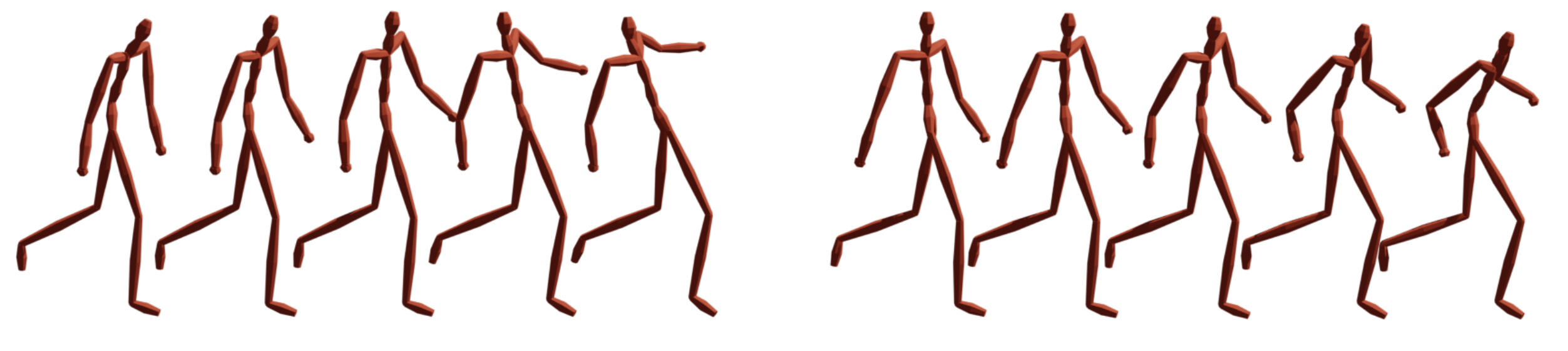} 
		\begin{tabular}{cc}
		(a) \hspace{3cm}&\hspace{3.6cm} (b) 
	\end{tabular}
	\caption{Style interpolation. Our style space code enables motion style interpolation. A neutral walking is transferred to an interpolated style. (a) Depressed to proud. (b) neutral to old.}
	\label{fig:interpolation}
\end{figure}

Our learned continuous style code space can be used to interpolate between styles. Style interpolation can be achieved by linearly interpolating between style code, and then decoding the results through our decoder. Our video demonstrates motions where the content input is fixed (neutral walking) and the style is interpolated between two different style codes (depressed to proud and neutral to old). Figure~\ref{fig:interpolation} shows a key-frame from each interpolated motion sequence.

\section{Conclusions and Future Work}
\label{sec:conc}

We have presented a neural network that transfers motion style from one sequence into another. 

The key novelty is that the network is trained without paired data, and without attempting to explicitly define either motion content or motion style. Nevertheless, the results show that the network succeeds to implicitly disentangle style and content and combine even previously unseen styles with a given content. We partly attribute the success of the approach to the asymmetric structure of the network, where the style is represented and controlled by instance normalization layers, while the content by deep convolutional layers. Instance normalization layers have the innate tendency to control mainly local statistics, or in other words, details, while the convolutional layers preserve the content.
Our training protocol, with various motion styles, encourages this asymmetric network to disentangle the latent style from the motion's content. 

Although there is no universally accepted definition of motion style, it may be argued that our framework defines style as the set of properties, shared by motions in a group, which can be manipulated (added/removed) by an affine, temporally invariant, transformation (AdaIN) applied to deep features. As a consequence, the complementary part of style that enables the reconstruction of local motion (without the root position), is defined as the content. In addition, the global positions, which are taken directly from the content input, are considered as part of the content, while the global velocity, which is temporally warped based on the style input, is defined to be part of the style in our case.

Our mechanism aims at disentangling style and content of arbitrary motions based on style labels. However, if the two input motions (content input and style input) during test time are different (lacking commonalities) and the target style departs too much from the ones used for training, the network will not be able to infer which style properties should be transferred. Moreover, due to the fact that the majority of the motions in our datasets depict locomotion (walking and running) the network tends to output motions of higher-quality with such samples during test time. In turn, this motivates us to use our generative system to produce more data and new styles by possibly mixing styles or amplifying (or attenuating) available styles or mixes of styles. 

Another notable limitation is that testing the system with characters that have different body proportions from those which were seen during training, may lead to implausible results.  In order to cope with such cases, motion retargeting should be performed prior to the style transfer pass. Motion retargeting is a challenging problem in its own right, and is outside the scope of this work. In order to support style transfer of various unseen skeletons in an end-to-end fashion, a different solution would have to be proposed. We leave this issue to future work.

In our current implementation, a given pair of input content and style motions yields a deterministic output.
We would like to consider extending the system by injecting noise to produce slight variations. This will allow a temporal prolongation of the input sequence without noticeable repetitions or discontinuous transitions. In the future, we would also consider segmenting the sequence temporally and transferring different styles to different segments.

We believe that the role of instance normalization in motion processing and animation is likely to increase, especially for generative models. The work we presented, is only a first step in that direction.  

\begin{acks}
We thank the anonymous reviewers for their constructive comments. This work was supported in part by  National Key R\&D Program of China (2018YFB1403900, 2019YFF0302902), and by the Israel Science Foundation (grant no.~2366/16).
\end{acks}

\bibliographystyle{ACM-Reference-Format}
\bibliography{main}


\begin{thebibliography}{43}


\ifx \showCODEN    \undefined \def \showCODEN     #1{\unskip}     \fi
\ifx \showDOI      \undefined \def \showDOI       #1{#1}\fi
\ifx \showISBNx    \undefined \def \showISBNx     #1{\unskip}     \fi
\ifx \showISBNxiii \undefined \def \showISBNxiii  #1{\unskip}     \fi
\ifx \showISSN     \undefined \def \showISSN      #1{\unskip}     \fi
\ifx \showLCCN     \undefined \def \showLCCN      #1{\unskip}     \fi
\ifx \shownote     \undefined \def \shownote      #1{#1}          \fi
\ifx \showarticletitle \undefined \def \showarticletitle #1{#1}   \fi
\ifx \showURL      \undefined \def \showURL       {\relax}        \fi
\providecommand\bibfield[2]{#2}
\providecommand\bibinfo[2]{#2}
\providecommand\natexlab[1]{#1}
\providecommand\showeprint[2][]{arXiv:#2}

\bibitem[\protect\citeauthoryear{Aberman, Shi, Liao, Lischinski, Chen, and
  Cohen-Or}{Aberman et~al\mbox{.}}{2019a}]%
        {aberman2019deep}
\bibfield{author}{\bibinfo{person}{Kfir Aberman}, \bibinfo{person}{Mingyi Shi},
  \bibinfo{person}{Jing Liao}, \bibinfo{person}{Dani Lischinski},
  \bibinfo{person}{Baoquan Chen}, {and} \bibinfo{person}{Daniel Cohen-Or}.}
  \bibinfo{year}{2019}\natexlab{a}.
\newblock \showarticletitle{Deep Video-Based Performance Cloning}. In
  \bibinfo{booktitle}{\emph{Computer Graphics Forum}},
  Vol.~\bibinfo{volume}{38}. Wiley Online Library, \bibinfo{pages}{219--233}.
\newblock


\bibitem[\protect\citeauthoryear{Aberman, Wu, Lischinski, Chen, and
  Cohen-Or}{Aberman et~al\mbox{.}}{2019b}]%
        {aberman2019learning}
\bibfield{author}{\bibinfo{person}{Kfir Aberman}, \bibinfo{person}{Rundi Wu},
  \bibinfo{person}{Dani Lischinski}, \bibinfo{person}{Baoquan Chen}, {and}
  \bibinfo{person}{Daniel Cohen-Or}.} \bibinfo{year}{2019}\natexlab{b}.
\newblock \showarticletitle{Learning Character-Agnostic Motion for Motion
  Retargeting in 2D}.
\newblock \bibinfo{journal}{\emph{ACM Transactions on Graphics (TOG)}}
  \bibinfo{volume}{38}, \bibinfo{number}{4} (\bibinfo{year}{2019}),
  \bibinfo{pages}{75}.
\newblock


\bibitem[\protect\citeauthoryear{Amaya, Bruderlin, and Calvert}{Amaya
  et~al\mbox{.}}{1996}]%
        {amaya1996emotion}
\bibfield{author}{\bibinfo{person}{Kenji Amaya}, \bibinfo{person}{Armin
  Bruderlin}, {and} \bibinfo{person}{Tom Calvert}.}
  \bibinfo{year}{1996}\natexlab{}.
\newblock \showarticletitle{Emotion from motion}. In
  \bibinfo{booktitle}{\emph{Graphics Interface}}, Vol.~\bibinfo{volume}{96}.
  Toronto, Canada, \bibinfo{pages}{222--229}.
\newblock


\bibitem[\protect\citeauthoryear{Aristidou, Cohen-Or, Hodgins, Chrysanthou, and
  Shamir}{Aristidou et~al\mbox{.}}{2018}]%
        {aristidou2018deep}
\bibfield{author}{\bibinfo{person}{Andreas Aristidou}, \bibinfo{person}{Daniel
  Cohen-Or}, \bibinfo{person}{Jessica~K. Hodgins}, \bibinfo{person}{Yiorgos
  Chrysanthou}, {and} \bibinfo{person}{Ariel Shamir}.}
  \bibinfo{year}{2018}\natexlab{}.
\newblock \showarticletitle{Deep Motifs and Motion Signatures}.
\newblock \bibinfo{journal}{\emph{ACM Trans. Graph.}} \bibinfo{volume}{37},
  \bibinfo{number}{6}, Article \bibinfo{articleno}{187} (\bibinfo{date}{Nov.}
  \bibinfo{year}{2018}), \bibinfo{numpages}{13}~pages.
\newblock
\urldef\tempurl%
\url{https://doi.org/10.1145/3272127.3275038}
\showDOI{\tempurl}


\bibitem[\protect\citeauthoryear{Aristidou, Zeng, Stavrakis, Yin, Cohen-Or,
  Chrysanthou, and Chen}{Aristidou et~al\mbox{.}}{2017}]%
        {aristidou2017emotion}
\bibfield{author}{\bibinfo{person}{Andreas Aristidou}, \bibinfo{person}{Qiong
  Zeng}, \bibinfo{person}{Efstathios Stavrakis}, \bibinfo{person}{KangKang
  Yin}, \bibinfo{person}{Daniel Cohen-Or}, \bibinfo{person}{Yiorgos
  Chrysanthou}, {and} \bibinfo{person}{Baoquan Chen}.}
  \bibinfo{year}{2017}\natexlab{}.
\newblock \showarticletitle{Emotion control of unstructured dance movements}.
  In \bibinfo{booktitle}{\emph{Proc.~ACM SIGGRAPH/Eurographics Symposium on
  Computer Animation}}. ACM, \bibinfo{pages}{9}.
\newblock


\bibitem[\protect\citeauthoryear{Brand and Hertzmann}{Brand and
  Hertzmann}{2000}]%
        {brand2000style}
\bibfield{author}{\bibinfo{person}{Matthew Brand} {and} \bibinfo{person}{Aaron
  Hertzmann}.} \bibinfo{year}{2000}\natexlab{}.
\newblock \showarticletitle{Style machines}. In
  \bibinfo{booktitle}{\emph{Proc.~SIGGRAPH 2000}}. ACM Press/Addison-Wesley
  Publishing Co., \bibinfo{pages}{183--192}.
\newblock


\bibitem[\protect\citeauthoryear{Cao, Hidalgo, Simon, Wei, and Sheikh}{Cao
  et~al\mbox{.}}{2018}]%
        {cao2018openpose}
\bibfield{author}{\bibinfo{person}{Zhe Cao}, \bibinfo{person}{Gines Hidalgo},
  \bibinfo{person}{Tomas Simon}, \bibinfo{person}{Shih-En Wei}, {and}
  \bibinfo{person}{Yaser Sheikh}.} \bibinfo{year}{2018}\natexlab{}.
\newblock \showarticletitle{OpenPose: realtime multi-person 2D pose estimation
  using Part Affinity Fields}.
\newblock \bibinfo{journal}{\emph{arXiv preprint arXiv:1812.08008}}
  (\bibinfo{year}{2018}).
\newblock


\bibitem[\protect\citeauthoryear{Chan, Ginosar, Zhou, and Efros}{Chan
  et~al\mbox{.}}{2019}]%
        {chan2019everybody}
\bibfield{author}{\bibinfo{person}{Caroline Chan}, \bibinfo{person}{Shiry
  Ginosar}, \bibinfo{person}{Tinghui Zhou}, {and} \bibinfo{person}{Alexei~A
  Efros}.} \bibinfo{year}{2019}\natexlab{}.
\newblock \showarticletitle{Everybody dance now}. In
  \bibinfo{booktitle}{\emph{Proceedings of the IEEE International Conference on
  Computer Vision}}. \bibinfo{pages}{5933--5942}.
\newblock


\bibitem[\protect\citeauthoryear{{CMU}}{{CMU}}{2019}]%
        {CMU:mocap}
\bibfield{author}{\bibinfo{person}{{CMU}}.} \bibinfo{year}{2019}\natexlab{}.
\newblock \bibinfo{title}{{CMU Graphics Lab Motion Capture Database}}.
\newblock
\newblock
\urldef\tempurl%
\url{http://mocap.cs.cmu.edu/}
\showURL{%
\tempurl}


\bibitem[\protect\citeauthoryear{Du, Herrmann, Sprenger, Cheema, Fischer,
  Slusallek, et~al\mbox{.}}{Du et~al\mbox{.}}{2019}]%
        {du2019stylistic}
\bibfield{author}{\bibinfo{person}{Han Du}, \bibinfo{person}{Erik Herrmann},
  \bibinfo{person}{Janis Sprenger}, \bibinfo{person}{Noshaba Cheema},
  \bibinfo{person}{Klaus Fischer}, \bibinfo{person}{Philipp Slusallek},
  {et~al\mbox{.}}} \bibinfo{year}{2019}\natexlab{}.
\newblock \showarticletitle{Stylistic Locomotion Modeling with Conditional
  Variational Autoencoder}. In \bibinfo{booktitle}{\emph{Proc.~Eurographics}}.
  \bibinfo{publisher}{The Eurographics Association}.
\newblock


\bibitem[\protect\citeauthoryear{Gatys, Ecker, and Bethge}{Gatys
  et~al\mbox{.}}{2016}]%
        {gatys2016image}
\bibfield{author}{\bibinfo{person}{Leon~A Gatys}, \bibinfo{person}{Alexander~S
  Ecker}, {and} \bibinfo{person}{Matthias Bethge}.}
  \bibinfo{year}{2016}\natexlab{}.
\newblock \showarticletitle{Image style transfer using convolutional neural
  networks}. In \bibinfo{booktitle}{\emph{Proc.~CVPR}}.
  \bibinfo{pages}{2414--2423}.
\newblock


\bibitem[\protect\citeauthoryear{Grochow, Martin, Hertzmann, and
  Popovi{\'c}}{Grochow et~al\mbox{.}}{2004}]%
        {grochow2004style}
\bibfield{author}{\bibinfo{person}{Keith Grochow}, \bibinfo{person}{Steven~L
  Martin}, \bibinfo{person}{Aaron Hertzmann}, {and} \bibinfo{person}{Zoran
  Popovi{\'c}}.} \bibinfo{year}{2004}\natexlab{}.
\newblock \showarticletitle{Style-based inverse kinematics}. In
  \bibinfo{booktitle}{\emph{ACM transactions on graphics (TOG)}},
  Vol.~\bibinfo{volume}{23}. ACM, \bibinfo{pages}{522--531}.
\newblock


\bibitem[\protect\citeauthoryear{Holden, Habibie, Kusajima, and Komura}{Holden
  et~al\mbox{.}}{2017a}]%
        {holden2017fast}
\bibfield{author}{\bibinfo{person}{Daniel Holden}, \bibinfo{person}{Ikhsanul
  Habibie}, \bibinfo{person}{Ikuo Kusajima}, {and} \bibinfo{person}{Taku
  Komura}.} \bibinfo{year}{2017}\natexlab{a}.
\newblock \showarticletitle{Fast neural style transfer for motion data}.
\newblock \bibinfo{journal}{\emph{IEEE computer graphics and applications}}
  \bibinfo{volume}{37}, \bibinfo{number}{4} (\bibinfo{year}{2017}),
  \bibinfo{pages}{42--49}.
\newblock


\bibitem[\protect\citeauthoryear{Holden, Komura, and Saito}{Holden
  et~al\mbox{.}}{2017b}]%
        {holden2017phase}
\bibfield{author}{\bibinfo{person}{Daniel Holden}, \bibinfo{person}{Taku
  Komura}, {and} \bibinfo{person}{Jun Saito}.}
  \bibinfo{year}{2017}\natexlab{b}.
\newblock \showarticletitle{Phase-functioned neural networks for character
  control}.
\newblock \bibinfo{journal}{\emph{ACM Transactions on Graphics (TOG)}}
  \bibinfo{volume}{36}, \bibinfo{number}{4} (\bibinfo{year}{2017}),
  \bibinfo{pages}{42}.
\newblock


\bibitem[\protect\citeauthoryear{Holden, Saito, and Komura}{Holden
  et~al\mbox{.}}{2016}]%
        {holden2016deep}
\bibfield{author}{\bibinfo{person}{Daniel Holden}, \bibinfo{person}{Jun Saito},
  {and} \bibinfo{person}{Taku Komura}.} \bibinfo{year}{2016}\natexlab{}.
\newblock \showarticletitle{A deep learning framework for character motion
  synthesis and editing}.
\newblock \bibinfo{journal}{\emph{ACM Transactions on Graphics (TOG)}}
  \bibinfo{volume}{35}, \bibinfo{number}{4} (\bibinfo{year}{2016}),
  \bibinfo{pages}{138}.
\newblock


\bibitem[\protect\citeauthoryear{Holden, Saito, Komura, and Joyce}{Holden
  et~al\mbox{.}}{2015}]%
        {holden2015learning}
\bibfield{author}{\bibinfo{person}{Daniel Holden}, \bibinfo{person}{Jun Saito},
  \bibinfo{person}{Taku Komura}, {and} \bibinfo{person}{Thomas Joyce}.}
  \bibinfo{year}{2015}\natexlab{}.
\newblock \showarticletitle{Learning motion manifolds with convolutional
  autoencoders}. In \bibinfo{booktitle}{\emph{SIGGRAPH Asia 2015 Technical
  Briefs}}. ACM, \bibinfo{pages}{18}.
\newblock


\bibitem[\protect\citeauthoryear{Hsu, Pulli, and Popovi{\'c}}{Hsu
  et~al\mbox{.}}{2005}]%
        {hsu2005style}
\bibfield{author}{\bibinfo{person}{Eugene Hsu}, \bibinfo{person}{Kari Pulli},
  {and} \bibinfo{person}{Jovan Popovi{\'c}}.} \bibinfo{year}{2005}\natexlab{}.
\newblock \showarticletitle{Style translation for human motion}. In
  \bibinfo{booktitle}{\emph{ACM Transactions on Graphics (TOG)}},
  Vol.~\bibinfo{volume}{24}. ACM, \bibinfo{pages}{1082--1089}.
\newblock


\bibitem[\protect\citeauthoryear{Huang and Belongie}{Huang and
  Belongie}{2017}]%
        {huang2017arbitrary}
\bibfield{author}{\bibinfo{person}{Xun Huang} {and} \bibinfo{person}{Serge
  Belongie}.} \bibinfo{year}{2017}\natexlab{}.
\newblock \showarticletitle{Arbitrary style transfer in real-time with adaptive
  instance normalization}. In \bibinfo{booktitle}{\emph{Proc.~ICCV}}.
  \bibinfo{pages}{1501--1510}.
\newblock


\bibitem[\protect\citeauthoryear{Huang, Liu, Belongie, and Kautz}{Huang
  et~al\mbox{.}}{2018}]%
        {huang2018multimodal}
\bibfield{author}{\bibinfo{person}{Xun Huang}, \bibinfo{person}{Ming-Yu Liu},
  \bibinfo{person}{Serge Belongie}, {and} \bibinfo{person}{Jan Kautz}.}
  \bibinfo{year}{2018}\natexlab{}.
\newblock \showarticletitle{Multimodal unsupervised image-to-image
  translation}. In \bibinfo{booktitle}{\emph{Proc.~ECCV}}.
  \bibinfo{pages}{172--189}.
\newblock


\bibitem[\protect\citeauthoryear{Ikemoto, Arikan, and Forsyth}{Ikemoto
  et~al\mbox{.}}{2009}]%
        {ikemoto2009generalizing}
\bibfield{author}{\bibinfo{person}{Leslie Ikemoto}, \bibinfo{person}{Okan
  Arikan}, {and} \bibinfo{person}{David Forsyth}.}
  \bibinfo{year}{2009}\natexlab{}.
\newblock \showarticletitle{Generalizing motion edits with gaussian processes}.
\newblock \bibinfo{journal}{\emph{ACM Transactions on Graphics (TOG)}}
  \bibinfo{volume}{28}, \bibinfo{number}{1} (\bibinfo{year}{2009}),
  \bibinfo{pages}{1}.
\newblock


\bibitem[\protect\citeauthoryear{Johnson, Alahi, and Fei-Fei}{Johnson
  et~al\mbox{.}}{2016}]%
        {johnson2016perceptual}
\bibfield{author}{\bibinfo{person}{Justin Johnson}, \bibinfo{person}{Alexandre
  Alahi}, {and} \bibinfo{person}{Li Fei-Fei}.} \bibinfo{year}{2016}\natexlab{}.
\newblock \showarticletitle{Perceptual losses for real-time style transfer and
  super-resolution}. In \bibinfo{booktitle}{\emph{Proc.~ECCV}}. Springer,
  \bibinfo{pages}{694--711}.
\newblock


\bibitem[\protect\citeauthoryear{Kanazawa, Zhang, Felsen, and Malik}{Kanazawa
  et~al\mbox{.}}{2019}]%
        {kanazawa2019learning}
\bibfield{author}{\bibinfo{person}{Angjoo Kanazawa}, \bibinfo{person}{Jason~Y
  Zhang}, \bibinfo{person}{Panna Felsen}, {and} \bibinfo{person}{Jitendra
  Malik}.} \bibinfo{year}{2019}\natexlab{}.
\newblock \showarticletitle{Learning 3d human dynamics from video}. In
  \bibinfo{booktitle}{\emph{Proceedings of the IEEE Conference on Computer
  Vision and Pattern Recognition}}. \bibinfo{pages}{5614--5623}.
\newblock


\bibitem[\protect\citeauthoryear{Karras, Laine, and Aila}{Karras
  et~al\mbox{.}}{2019}]%
        {karras2019style}
\bibfield{author}{\bibinfo{person}{Tero Karras}, \bibinfo{person}{Samuli
  Laine}, {and} \bibinfo{person}{Timo Aila}.} \bibinfo{year}{2019}\natexlab{}.
\newblock \showarticletitle{A style-based generator architecture for generative
  adversarial networks}. In \bibinfo{booktitle}{\emph{Proc.~CVPR}}.
  \bibinfo{pages}{4401--4410}.
\newblock


\bibitem[\protect\citeauthoryear{Liu, Hertzmann, and Popovi{\'c}}{Liu
  et~al\mbox{.}}{2005}]%
        {liu2005learning}
\bibfield{author}{\bibinfo{person}{C~Karen Liu}, \bibinfo{person}{Aaron
  Hertzmann}, {and} \bibinfo{person}{Zoran Popovi{\'c}}.}
  \bibinfo{year}{2005}\natexlab{}.
\newblock \showarticletitle{Learning physics-based motion style with nonlinear
  inverse optimization}. In \bibinfo{booktitle}{\emph{ACM Transactions on
  Graphics (TOG)}}, Vol.~\bibinfo{volume}{24}. ACM,
  \bibinfo{pages}{1071--1081}.
\newblock


\bibitem[\protect\citeauthoryear{Liu, Xu, Zollhoefer, Kim, Bernard, Habermann,
  Wang, and Theobalt}{Liu et~al\mbox{.}}{2018}]%
        {liu2018neural}
\bibfield{author}{\bibinfo{person}{Lingjie Liu}, \bibinfo{person}{Weipeng Xu},
  \bibinfo{person}{Michael Zollhoefer}, \bibinfo{person}{Hyeongwoo Kim},
  \bibinfo{person}{Florian Bernard}, \bibinfo{person}{Marc Habermann},
  \bibinfo{person}{Wenping Wang}, {and} \bibinfo{person}{Christian Theobalt}.}
  \bibinfo{year}{2018}\natexlab{}.
\newblock \showarticletitle{Neural Rendering and Reenactment of Human Actor
  Videos}.
\newblock \bibinfo{journal}{\emph{arXiv preprint arXiv:1809.03658}}
  (\bibinfo{year}{2018}).
\newblock


\bibitem[\protect\citeauthoryear{Liu, Huang, Mallya, Karras, Aila, Lehtinen,
  and Kautz}{Liu et~al\mbox{.}}{2019}]%
        {liu2019few}
\bibfield{author}{\bibinfo{person}{Ming-Yu Liu}, \bibinfo{person}{Xun Huang},
  \bibinfo{person}{Arun Mallya}, \bibinfo{person}{Tero Karras},
  \bibinfo{person}{Timo Aila}, \bibinfo{person}{Jaakko Lehtinen}, {and}
  \bibinfo{person}{Jan Kautz}.} \bibinfo{year}{2019}\natexlab{}.
\newblock \showarticletitle{Few-shot unsupervised image-to-image translation}.
\newblock \bibinfo{journal}{\emph{arXiv preprint arXiv:1905.01723}}
  (\bibinfo{year}{2019}).
\newblock


\bibitem[\protect\citeauthoryear{Ma, Xia, Hodgins, Yang, Li, and Wang}{Ma
  et~al\mbox{.}}{2010}]%
        {ma2010modeling}
\bibfield{author}{\bibinfo{person}{Wanli Ma}, \bibinfo{person}{Shihong Xia},
  \bibinfo{person}{Jessica~K Hodgins}, \bibinfo{person}{Xiao Yang},
  \bibinfo{person}{Chunpeng Li}, {and} \bibinfo{person}{Zhaoqi Wang}.}
  \bibinfo{year}{2010}\natexlab{}.
\newblock \showarticletitle{Modeling style and variation in human motion}. In
  \bibinfo{booktitle}{\emph{Proc.~2010 ACM SIGGRAPH/Eurographics Symposium on
  Computer Animation}}. Eurographics Association, \bibinfo{pages}{21--30}.
\newblock


\bibitem[\protect\citeauthoryear{Mason, Starke, Zhang, Bilen, and Komura}{Mason
  et~al\mbox{.}}{2018}]%
        {mason2018few}
\bibfield{author}{\bibinfo{person}{Ian Mason}, \bibinfo{person}{Sebastian
  Starke}, \bibinfo{person}{He Zhang}, \bibinfo{person}{Hakan Bilen}, {and}
  \bibinfo{person}{Taku Komura}.} \bibinfo{year}{2018}\natexlab{}.
\newblock \showarticletitle{Few-shot Learning of Homogeneous Human Locomotion
  Styles}. In \bibinfo{booktitle}{\emph{Computer Graphics Forum}},
  Vol.~\bibinfo{volume}{37}. Wiley Online Library, \bibinfo{pages}{143--153}.
\newblock


\bibitem[\protect\citeauthoryear{Mehta, Sridhar, Sotnychenko, Rhodin, Shafiei,
  Seidel, Xu, Casas, and Theobalt}{Mehta et~al\mbox{.}}{2017}]%
        {Mehta:2017}
\bibfield{author}{\bibinfo{person}{Dushyant Mehta}, \bibinfo{person}{Srinath
  Sridhar}, \bibinfo{person}{Oleksandr Sotnychenko}, \bibinfo{person}{Helge
  Rhodin}, \bibinfo{person}{Mohammad Shafiei}, \bibinfo{person}{Hans-Peter
  Seidel}, \bibinfo{person}{Weipeng Xu}, \bibinfo{person}{Dan Casas}, {and}
  \bibinfo{person}{Christian Theobalt}.} \bibinfo{year}{2017}\natexlab{}.
\newblock \showarticletitle{VNect: Real-time 3D Human Pose Estimation with a
  Single RGB Camera}.
\newblock \bibinfo{journal}{\emph{ACM Trans. Graph.}} \bibinfo{volume}{36},
  \bibinfo{number}{4}, Article \bibinfo{articleno}{44} (\bibinfo{date}{July}
  \bibinfo{year}{2017}), \bibinfo{numpages}{44:1--44:14}~pages.
\newblock
\showISSN{0730-0301}
\urldef\tempurl%
\url{https://doi.org/10.1145/3072959.3073596}
\showDOI{\tempurl}


\bibitem[\protect\citeauthoryear{Park, Liu, Wang, and Zhu}{Park
  et~al\mbox{.}}{2019}]%
        {park2019semantic}
\bibfield{author}{\bibinfo{person}{Taesung Park}, \bibinfo{person}{Ming-Yu
  Liu}, \bibinfo{person}{Ting-Chun Wang}, {and} \bibinfo{person}{Jun-Yan Zhu}.}
  \bibinfo{year}{2019}\natexlab{}.
\newblock \showarticletitle{Semantic image synthesis with spatially-adaptive
  normalization}. In \bibinfo{booktitle}{\emph{Proc.~CVPR}}.
  \bibinfo{pages}{2337--2346}.
\newblock


\bibitem[\protect\citeauthoryear{Pavllo, Feichtenhofer, Auli, and
  Grangier}{Pavllo et~al\mbox{.}}{2019a}]%
        {pavllo2019modeling}
\bibfield{author}{\bibinfo{person}{Dario Pavllo}, \bibinfo{person}{Christoph
  Feichtenhofer}, \bibinfo{person}{Michael Auli}, {and} \bibinfo{person}{David
  Grangier}.} \bibinfo{year}{2019}\natexlab{a}.
\newblock \showarticletitle{Modeling Human Motion with Quaternion-based Neural
  Networks}.
\newblock \bibinfo{journal}{\emph{arXiv preprint arXiv:1901.07677}}
  (\bibinfo{year}{2019}).
\newblock


\bibitem[\protect\citeauthoryear{Pavllo, Feichtenhofer, Grangier, and
  Auli}{Pavllo et~al\mbox{.}}{2019b}]%
        {pavllo20193d}
\bibfield{author}{\bibinfo{person}{Dario Pavllo}, \bibinfo{person}{Christoph
  Feichtenhofer}, \bibinfo{person}{David Grangier}, {and}
  \bibinfo{person}{Michael Auli}.} \bibinfo{year}{2019}\natexlab{b}.
\newblock \showarticletitle{3D human pose estimation in video with temporal
  convolutions and semi-supervised training}. In
  \bibinfo{booktitle}{\emph{Proceedings of the IEEE Conference on Computer
  Vision and Pattern Recognition}}. \bibinfo{pages}{7753--7762}.
\newblock


\bibitem[\protect\citeauthoryear{Shapiro, Cao, and Faloutsos}{Shapiro
  et~al\mbox{.}}{2006}]%
        {shapiro2006style}
\bibfield{author}{\bibinfo{person}{Ari Shapiro}, \bibinfo{person}{Yong Cao},
  {and} \bibinfo{person}{Petros Faloutsos}.} \bibinfo{year}{2006}\natexlab{}.
\newblock \showarticletitle{Style components}. In
  \bibinfo{booktitle}{\emph{Proc.~Graphics Interface 2006}}. Canadian
  Information Processing Society, \bibinfo{pages}{33--39}.
\newblock


\bibitem[\protect\citeauthoryear{Smith, Cao, Neff, and Wang}{Smith
  et~al\mbox{.}}{2019}]%
        {smith2019efficient}
\bibfield{author}{\bibinfo{person}{Harrison~Jesse Smith}, \bibinfo{person}{Chen
  Cao}, \bibinfo{person}{Michael Neff}, {and} \bibinfo{person}{Yingying Wang}.}
  \bibinfo{year}{2019}\natexlab{}.
\newblock \showarticletitle{Efficient Neural Networks for Real-time Motion
  Style Transfer}.
\newblock \bibinfo{journal}{\emph{PACMCGIT}} \bibinfo{volume}{2},
  \bibinfo{number}{2} (\bibinfo{year}{2019}), \bibinfo{pages}{13:1--13:17}.
\newblock


\bibitem[\protect\citeauthoryear{Taylor and Hinton}{Taylor and Hinton}{2009}]%
        {taylor2009factored}
\bibfield{author}{\bibinfo{person}{Graham~W Taylor} {and}
  \bibinfo{person}{Geoffrey~E Hinton}.} \bibinfo{year}{2009}\natexlab{}.
\newblock \showarticletitle{Factored conditional restricted Boltzmann machines
  for modeling motion style}. In \bibinfo{booktitle}{\emph{Proc.~ICML}}. ACM,
  \bibinfo{pages}{1025--1032}.
\newblock


\bibitem[\protect\citeauthoryear{Ulyanov, Vedaldi, and Lempitsky}{Ulyanov
  et~al\mbox{.}}{2016}]%
        {ulyanov2016instance}
\bibfield{author}{\bibinfo{person}{Dmitry Ulyanov}, \bibinfo{person}{Andrea
  Vedaldi}, {and} \bibinfo{person}{Victor Lempitsky}.}
  \bibinfo{year}{2016}\natexlab{}.
\newblock \showarticletitle{Instance normalization: The missing ingredient for
  fast stylization}.
\newblock \bibinfo{journal}{\emph{arXiv preprint arXiv:1607.08022}}
  (\bibinfo{year}{2016}).
\newblock


\bibitem[\protect\citeauthoryear{Unuma, Anjyo, and Takeuchi}{Unuma
  et~al\mbox{.}}{1995}]%
        {unuma1995fourier}
\bibfield{author}{\bibinfo{person}{Munetoshi Unuma}, \bibinfo{person}{Ken
  Anjyo}, {and} \bibinfo{person}{Ryozo Takeuchi}.}
  \bibinfo{year}{1995}\natexlab{}.
\newblock \showarticletitle{Fourier principles for emotion-based human figure
  animation}. In \bibinfo{booktitle}{\emph{Proc.~SIGGRAPH '95}}. ACM,
  \bibinfo{pages}{91--96}.
\newblock


\bibitem[\protect\citeauthoryear{Villegas, Yang, Ceylan, and Lee}{Villegas
  et~al\mbox{.}}{2018}]%
        {villegas2018neural}
\bibfield{author}{\bibinfo{person}{Ruben Villegas}, \bibinfo{person}{Jimei
  Yang}, \bibinfo{person}{Duygu Ceylan}, {and} \bibinfo{person}{Honglak Lee}.}
  \bibinfo{year}{2018}\natexlab{}.
\newblock \showarticletitle{Neural kinematic networks for unsupervised motion
  retargetting}. In \bibinfo{booktitle}{\emph{Proceedings of the IEEE
  Conference on Computer Vision and Pattern Recognition}}.
  \bibinfo{pages}{8639--8648}.
\newblock


\bibitem[\protect\citeauthoryear{Wang, Fleet, and Hertzmann}{Wang
  et~al\mbox{.}}{2007}]%
        {wang2007multifactor}
\bibfield{author}{\bibinfo{person}{Jack~M Wang}, \bibinfo{person}{David~J
  Fleet}, {and} \bibinfo{person}{Aaron Hertzmann}.}
  \bibinfo{year}{2007}\natexlab{}.
\newblock \showarticletitle{Multifactor Gaussian process models for
  style-content separation}. In \bibinfo{booktitle}{\emph{Proc.~ICML}}. ACM,
  \bibinfo{pages}{975--982}.
\newblock


\bibitem[\protect\citeauthoryear{Wang, Liu, Zhu, Tao, Kautz, and
  Catanzaro}{Wang et~al\mbox{.}}{2018}]%
        {wang2018high}
\bibfield{author}{\bibinfo{person}{Ting-Chun Wang}, \bibinfo{person}{Ming-Yu
  Liu}, \bibinfo{person}{Jun-Yan Zhu}, \bibinfo{person}{Andrew Tao},
  \bibinfo{person}{Jan Kautz}, {and} \bibinfo{person}{Bryan Catanzaro}.}
  \bibinfo{year}{2018}\natexlab{}.
\newblock \showarticletitle{High-resolution image synthesis and semantic
  manipulation with conditional gans}. In \bibinfo{booktitle}{\emph{Proceedings
  of the IEEE conference on computer vision and pattern recognition}}.
  \bibinfo{pages}{8798--8807}.
\newblock


\bibitem[\protect\citeauthoryear{Xia, Wang, Chai, and Hodgins}{Xia
  et~al\mbox{.}}{2015}]%
        {xia2015realtime}
\bibfield{author}{\bibinfo{person}{Shihong Xia}, \bibinfo{person}{Congyi Wang},
  \bibinfo{person}{Jinxiang Chai}, {and} \bibinfo{person}{Jessica Hodgins}.}
  \bibinfo{year}{2015}\natexlab{}.
\newblock \showarticletitle{Realtime style transfer for unlabeled heterogeneous
  human motion}.
\newblock \bibinfo{journal}{\emph{ACM Transactions on Graphics (TOG)}}
  \bibinfo{volume}{34}, \bibinfo{number}{4} (\bibinfo{year}{2015}),
  \bibinfo{pages}{119}.
\newblock


\bibitem[\protect\citeauthoryear{Yumer and Mitra}{Yumer and Mitra}{2016}]%
        {yumer2016spectral}
\bibfield{author}{\bibinfo{person}{M~Ersin Yumer} {and}
  \bibinfo{person}{Niloy~J Mitra}.} \bibinfo{year}{2016}\natexlab{}.
\newblock \showarticletitle{Spectral style transfer for human motion between
  independent actions}.
\newblock \bibinfo{journal}{\emph{ACM Transactions on Graphics (TOG)}}
  \bibinfo{volume}{35}, \bibinfo{number}{4} (\bibinfo{year}{2016}),
  \bibinfo{pages}{137}.
\newblock


\bibitem[\protect\citeauthoryear{Zhou, Barnes, Lu, Yang, and Li}{Zhou
  et~al\mbox{.}}{2019}]%
        {zhou2019continuity}
\bibfield{author}{\bibinfo{person}{Yi Zhou}, \bibinfo{person}{Connelly Barnes},
  \bibinfo{person}{Jingwan Lu}, \bibinfo{person}{Jimei Yang}, {and}
  \bibinfo{person}{Hao Li}.} \bibinfo{year}{2019}\natexlab{}.
\newblock \showarticletitle{On the continuity of rotation representations in
  neural networks}. In \bibinfo{booktitle}{\emph{Proceedings of the IEEE
  Conference on Computer Vision and Pattern Recognition}}.
  \bibinfo{pages}{5745--5753}.
\newblock


\end{thebibliography}

\end{document}